\documentclass[useAMS,usenatbib]{mn2e}
\usepackage{epsfig,rotate,graphicx}
\usepackage[fleqn]{amsmath}
\usepackage{subfigure}
\usepackage{lscape}
\usepackage{bm}
\usepackage{paralist}

\newcommand{\p}{\partial}
\newcommand{\mnras}{MNRAS}
\newcommand{\apj}{ApJ}
\newcommand{\na}{New A}
\newcommand{\aap}{A\&A}
\newcommand{\apjl}{ApJL}
\newcommand{\araa}{ARAA}

\newcommand{\be}{\begin{equation}}
\newcommand{\ee}{\end{equation}}
\newcommand{\gtrsim}{\;\raisebox{-.8ex}{$\buildrel{\textstyle>}\over\sim$}\;}
\newcommand{\lesssim}{\; \raisebox{-.8ex}{$\buildrel{\textstyle<}\over\sim$}\;}

\newcommand{\apjs}{{\it ApJS, }}

\newcommand{\aaps}{{\it A\&AS, }}

\newcommand{\avg}[1]{\langle #1 \rangle}
\newcommand{\aziavg}[1]{\langle #1 \rangle_\phi}

\newcommand{\bv}{\bm{v}}
\newcommand{\rin}{r_\mathrm{in}}
\newcommand{\rout}{r_\mathrm{out}}
\newcommand{\dampin}{r_\mathrm{d,in}}
\newcommand{\dampout}{r_\mathrm{d,out}}

\newcommand{\psimax}{\psi_\mathrm{max}}

\newcommand{\pluto}{{\tt PLUTO }}
\newcommand{\ii}{\mathrm{i}}

\title[Vortices in viscous discs]{Testing large-scale 
  vortex formation against viscous layers in three-dimensional discs}

\author[Lin]{Min-Kai Lin
  \thanks{E-mail: mklin924@cita.utoronto.ca} \\
Canadian Institute for Theoretical Astrophysics,
  60 St. George Street, Toronto, ON, M5S 3H8, Canada 
}

\begin{document}

\maketitle
\begin{abstract}
  Vortex formation through the Rossby wave 
  instability (RWI) in protoplanetary discs has been invoked to play a
  role in planet formation theory, and suggested to explain
  the observation of large dust asymmetries in  several transitional discs.   
  However, whether or not vortex formation operates in layered
  accretion discs, i.e. models of protoplanetary discs including dead
  zones  near the disc midplane --- regions that are magnetically
    inactive and the effective viscosity greatly reduced --- 
  has not been verified. As a first step toward testing the robustness of
  vortex formation in layered discs, we present 
  non-linear hydrodynamical 
  simulations of global 3D protoplanetary discs with an imposed kinematic
  viscosity that increases away from the disc midplane. Two sets of numerical 
  experiments are performed:
  \begin{inparaenum}[(i)]
  \item non-axisymmetric instability of artificial
     radial density bumps in viscous discs;\label{exp1}  
  \item vortex-formation at planetary gap edges in layered discs.
  \end{inparaenum}\,
   Experiment (\ref{exp1}) shows that the linear instability is
   largely unaffected by viscosity and remains dynamical.   
   The disc-planet simulations  
   also show the initial development of vortices at gap edges, but in
   layered discs the vortices are transient structures which disappear  
    well into the non-linear regime. We suggest that  the long term survival of 
     columnar vortices, such as those formed via the RWI, 
     requires low effective viscosity throughout the vertical extent
     of the disc, so such vortices do not persist in layered discs.  
   
%
\end{abstract}

\begin{keywords}
planetary systems: formation --- planetary systems:
protoplanetary discs
\end{keywords}

\section{Introduction}\label{intro}


Recent observations have revealed a class of transition discs ---
circumstellar discs which are dust poor in its inner regions ---  
with non-axisymmetric dust distributions in its outer 
regions \citep{brown09,mayama12,marel13,isella13}.  
One interpretation of such a non-axisymmetric structure is the presence of
a large-scale disc vortex, which is known to act as a dust trap
\citep{barge95,inaba06,birnstiel13,ataiee13,lyra13}.  
Because of its occurrence adjacent to the inner dust
hole, i.e. a cavity edge, it has been suggested that such a vortex is
a result of the Rossby wave instability (RWI): a hydrodynamical instability
that can develop in radially structured discs. 

Modern work on the RWI began with two-dimensional (2D) linear
stability analysis \citep{lovelace99,li00}. These studies show that a
disc with radially localized 
structure, such as a surface density enhancement of $\gtrsim 10\%$ over a
radial length scale of order the local disc scale-height, is unstable to
non-axisymmetric perturbations, which grow on dynamical (orbital)
timescales. Early 2D non-linear hydrodynamic simulations showed that
the RWI leads to multi-vortex formation, followed by vortex merging into
a single large vortex in quasi-steady state \citep{li01,inaba06}. 

While these studies consider disc models with artificial radial
structure, it has recently been established that a natural site for
the RWI is the edge of gaps induced by disc-planet interaction 
\citep{koller03,li05,valborro07,li09,lyra09b,lin10,lin11a}. Indeed, this has
been the proposed explanation for the lopsided dust distribution
observed in the Oph IRS 48 transition disc system
\citep{marel13}. 

An important extension to the aforementioned studies is
the generalization of the RWI to three-dimensional (3D) 
discs. Both non-linear 3D hydrodynamic simulations 
\citep{meheut10,meheut12b,lin12b,lyra12,richard13b} and 3D linear stability calculations
\citep{umurhan10,meheut12,lin12,lin13} have been carried out. 
These studies reveal that the RWI is a 2D instability,
in that there is negligible difference between growth rates obtained
from 2D and 3D linear calculations. The associated density and
horizontal velocity perturbations have weak vertical dependence; and  
vertical velocities are small. In non-linear hydrodynamic simulations,
the vortices are columnar and extend throughout the vertical extent of
the disc \citep{richard13b}.  

The RWI therefore appears to be a global instability in the direction
perpendicular to the disc midplane: the vortical perturbation involves
the entire fluid column. Thus conditions away
from the disc midplane may have important effects on vortex formation
via the RWI. For example, \cite{lin13a} only found linear
instability for certain upper disc boundary conditions. This issue is
relevant to protoplanetary disc models including `dead zones'.   

It is believed that mass accretion in protoplanetary discs is driven by
magneto-hydrodynamic (MHD) turbulence as a result of the
magneto-rotational instability \citep[MRI,][]{balbus91,balbus98}. However, it
is not clear if the MRI operates throughout the vertical extent of the
disc, because the midplane of protoplanetary 
disc is dense and cold \citep{armitage11}. As a result, \cite{gammie96}
proposed the layered disc model: accretion due to MHD turbulence
is small near the midplane (the dead zone), while MHD
turbulence-driven accretion  
operates near the disc surface (the active zone). The layered
accretion disc model has been subject to numerous studies 
\citep[e.g.,][]{fleming03,terquem08,oishi09,dzy10,kretke10,okuzumi11,flaig11,landry13}.      
If MRI-driven accretion can be modeled through an effective 
viscosity \citep{balbus99}, this corresponds to a low viscosity midplane and 
high viscosity atmosphere. It is therefore valid to ask how such a
vertical disc structure would affect large-scale vortex formation via
the RWI.  

This problem is partly motivated by viscous disc-planet
simulations which show that  gap-edge vortex formation only
occurs when the viscosity is sufficiently small
\citep{valborro06,valborro07,edgar08}. 
What happens if the effective viscosity near the midplane is
sufficiently low for the development of Rossby vortices, 
but is too high away from the midplane? 

In this work we examine vortex formation through the RWI in
layered discs. As a first study, we take an
experimental approach through customized numerical hydrodynamic
simulations. We simulate global 3D protoplanetary discs with an imposed
kinematic viscosity that varies with height above the disc
midplane. 
The central question is whether or not applying a viscosity only in the upper
layers of the disc damps the RWI and subsequent vortex formation.   
The purpose of this paper is to demonstrate, through selected 
simulations, the potential importance of layered disc structures on
vortex formation. We defer a detailed parameter survey to a future
study.

This paper is organised as follows. The accretion disc model is 
set up in \S\ref{model} and the numerical simulation method described
in \S\ref{sims}. Results are presented in \S\ref{density_bump} for 
viscous discs initialised with a density bump. These 
simulations employ a special setup such that the density bump is not
subject to axisymmetric viscous diffusion. This allows one to
focus on the effect of layered viscosity on the linear
non-axisymmetric instability. \S\ref{disc-planet} revisits vortex
formation at planetary gap edges, but in 3D layered discs, 
where it will be seen that vortex formation can be suppressed by
viscous layers. \S\ref{summary} concludes this work with a discussion
of important caveats of the present disc models.   

\section{Disc model and Governing equations}\label{model} 
We consider a three-dimensional,  locally isothermal,
non-self-gravitating fluid disc orbiting a central 
star of mass $M_*$.  We adopt a non-rotating frame centred on the
  star. Our computer simulations employ spherical co-ordinates
  $\bm{r}=(r,\theta,\phi)$, but for model description and results 
  analysis we will also use cylindrical co-ordinates $\bm{r}=(R, \phi,
  z)$. We also define $\psi \equiv \pi/2 -
\theta$ as the angular displacement from the disc midplane. For
convenience, we will sometimes refer to $\psi$ as the vertical
direction. The governing equations are: 
\begin{align}
  &\frac{\p\rho}{\p t} + \nabla\cdot\left(\rho\bm{v}\right) =
  0,\label{cont_eq}\\ 
  &\frac{\p\bm{v}}{\p t} + \bm{v}\cdot\nabla\bv = -\frac{1}{\rho}\nabla
  p -\nabla{\left(\Phi_*+\Phi_p\right)} +\bm{f}_\nu + \bm{f}_d\label{mom_eq},    
\end{align}
where $\rho$ is the mass density, $\bv$ is the velocity field (the
azimuthal angular velocity being $\Omega\equiv v_\phi/R$) and 
$p=c_s^2\rho$ is the pressure. 
The sound speed $c_s$ is prescribed as 
\begin{align}
  c_s = hr_0\Omega_k(r_0)\times\left(\frac{r_0}{R}\right)^{q/2}, 
\end{align}
where $h$ is the aspect-ratio at the reference radius $r_0$, 
$\Omega_k(R) = \sqrt{GM_*/R^3}$ is the Keplerian frequency and $G$ is
the gravitational constant. The power-law index $q$ specifies the
radial temperature profile: $q=0$ corresponds to a strictly isothermal
disc, while $q=1$ is a locally isothermal disc with constant aspect
ratio. In Eq. \ref{mom_eq}, $\Phi_*(r) = -GM_*/r $ is the stellar
potential and $\Phi_p$ is a planetary potential (see \S\ref{planet}
for details).  

Two dissipative terms are included in the momentum, equations: viscous
damping $\bm{f}_\nu$ and frictional damping $\bm{f}_d$. The viscous
force is 
\begin{align}
  \bm{f}_\nu = \frac{1}{\rho}\nabla\cdot\bm{T},
\end{align}
where 
\begin{align}
  \bm{T} = \rho\nu \left[\nabla\bv + \left(\nabla\bv\right)^\dagger
    - \frac{2}{3}\left(\nabla\cdot\bv\right)\bm{1} \right]
\end{align}
is the viscous stress tensor and $\nu$ is the kinematic viscosity 
($^\dagger$ denotes the transpose). The frictional force is 
\begin{align}
  \bm{f}_d = -\gamma\left(\bv - \bv_\mathrm{ref}\right),
\end{align}
where $\gamma$ is the damping coefficient and
$\bm{v}_\mathrm{ref}$ is a reference velocity field. 
$\nu$ and $\gamma$ are prescribed functions of position only (see
below). 

\subsection{Disc model and initial conditions}\label{IC}
The numerical disc model occupies $r\in[\rin,\rout]$,
$\theta\in[\theta_\mathrm{min}, 
  \pi/2]$ and $\phi\in[0,2\pi]$ in spherical co-ordinates. 
Only the upper disc is simulated explicitly 
($\psi>0$), by assuming symmetry across the midplane. The maximum
angular height is $\psimax \equiv \pi/2 - \theta_\mathrm{min}$. 
The extent of the vertical domain is parametrized by $n_h\equiv
\tan{\psimax}/h$, i.e. the number of scale-heights at the reference 
radius.   

The disc is initially axisymmetric with zero cylindrical vertical
velocity: $\rho(t=0)\equiv\rho_i(R,z)$ and
$\bm{v}(t=0)\equiv(v_{Ri}\,, R\Omega_i\,, 0)$ in cylindrical
co-ordinates. The initial density field is set by assuming vertical
hydrostatic balance between gas pressure and stellar gravity:
\begin{align}\label{vert_balance}
  0 = \frac{1}{\rho_i}\frac{\p p_i}{\p z} + \frac{\p \Phi_*}{\p z},
\end{align}
where $p_i=c_s^2\rho_i$ is the initial pressure field. We write 
\begin{align}\label{init_den}
  \rho_i = \frac{\Sigma_i(R)}{\sqrt{2\pi}H(R)}
  \exp{\left\{\frac{1}{c_s^2}\left[\Phi_*(R) - \Phi_*(r)\right]\right\}},
\end{align}
where $H = c_s/\Omega_k$ is the pressure scale-height. The initial
surface density $\Sigma_i(R)$ is chosen as  
\begin{align}
  \Sigma_i(R) = \Sigma_0\left(\frac{R}{r_0}\right)^{-\sigma}\times B(R),
\end{align}
where $\sigma$ is the power-law index, and the surface density scale $\Sigma_0$ 
is arbitrary for a non-self-gravitating disc. 
The bump function $B(R)$ is 
\begin{align}\label{bump_func}
B(R) = 1 + \left(A - 1\right)\exp{\left[-\frac{(R-r_0)^2}{2\Delta R^2}\right]},
\end{align}
where $A$ is the bump amplitude and $\Delta R$ is the bump width. The
initial surface  
density has bump if $A>1$ and is smooth if $A=1$. 

The initial angular velocity is chosen to 
satisfy centrifugal balance with pressure and stellar gravity: 
\begin{align}\label{init_vphi} 
  R\Omega^2_i = \frac{1}{\rho_i}\frac{\p p_i}{\p
    R} + \frac{\p\Phi_*}{\p R},  
\end{align}
so $\Omega_i=\Omega_i(R)$ for a strictly isothermal equation of
state ($q=0$). 

The initial cylindrical radial velocity $v_{Ri}$ and the viscosity
profile $\nu$ depends on the numerical experiment, and will be
described along with simulation results. Note that $v_{Ri}$ and $\nu$
are not independent if one additionally requires a steady-state (see
\S\ref{density_bump}).    

\subsection{Damping}
We apply frictional damping in the radial direction to reduce
reflections from boundaries \citep[e.g.][]{bate02,valborro07}.  The
damping coefficient $\gamma$ is only  non-zero within the `damping
zones', here taken to be $r\leq \dampin,\,  
r\geq \dampout$, 
\begin{align}
  \gamma = \hat{\gamma}\Omega_i\times
  \begin{cases}
    \xi_\mathrm{in}(r) & r\leq\dampin \\
    \xi_\mathrm{out}(r) & r\geq\dampout \\
  \end{cases},
\end{align}
where $\hat{\gamma}$ is the dimensionless damping rate. We choose
\begin{align}
  \xi_\mathrm{in}(r) = \left(\frac{\dampin - r}{\dampin - \rin}\right)^2 \text{ and } \quad
  \xi_\mathrm{out}(r) = \left(\frac{r - \dampout}{\rout- \dampout}\right)^2
\end{align}
for the inner and outer radial zones, respectively. 

\subsection{Planet potential}\label{planet}
Our disc model has the option to include a planet potential $\Phi_p$,
\begin{align}
  \Phi_p(\bm{r},t) = -\frac{GM_p}{\sqrt{|\bm{r}-\bm{r}_p(t)|^2 +
      \epsilon^2_p}} + \frac{GM_p}{|\bm{r}_p|^3}\bm{r}\cdot\bm{r}_p,
\end{align}
where $M_p$ is the planet mass,
$\bm{r}_p(t)~=~(r_0,\,\pi/2,\,\Omega_k(r_0)t+\pi)$ its position in
spherical co-ordinates, $\epsilon_p = \epsilon_{p0}r_h$ is a
softening length, and $r_h=(M_p/3M_*)^{1/3}r_0$ is the Hill radius. 
For the purpose of our study $\Phi_p$ is considered as a 
fixed external potential. That is, orbital migration is neglected. 

\section{Numerical experiments}\label{sims}
The necessary condition for the RWI --- a potential vorticity 
extremum \citep{li00} --- is either set as an initial condition
via a density bump, or obtained from a smooth disc by evolving it
under disc-planet interaction. The setup of each experiment is
detailed in subsequent sections.    

We adopt units such that $G=M_*=1$, and the reference radius $r_0=1$.    
We set $\sigma=0.5$ for the initial surface density profile and apply 
frictional damping within the shells $r<r_\mathrm{d,in} = 1.25\rin$
and $r>r_\mathrm{d,out}=0.84\rout$.

The fluid equations are evolved using the \pluto code \citep{mignone07} with 
the FARGO algorithm enabled  \citep{masset00,mignone12}. We employ a static
spherical grid with $(N_r, N_\theta, N_\phi)$ zones uniformly spaced
in all directions. For the present simulations the code was configured
with piece-wise linear reconstruction, a Roe solver and second order
Runge-Kutta time integration.   

Boundary conditions are imposed through ghost zones.   
Let the flow velocity parallel and normal to a boundary be
$v_\parallel$ and $v_\perp$, respectively. Two types of numerical 
conditions are considered for the $(r,\theta)$ boundaries:
\begin{inparaenum}[(a)]
\item \emph{reflective}: $\rho$ and $v_\parallel$ are symmetric with
  respect to the boundary while $v_\perp$ is anti-symmetric;  
\item \emph{unperturbed}: ghost zones retain their initial values
\end{inparaenum}. 
The boundary conditions adopted for all simulations is unperturbed in
$r$, reflective in $\theta$ and periodic in $\phi$.

\subsection{Diagnostics}
We list several quantities calculated from simulation data for use in
results visualization and analysis.  
 
\subsubsection{Density perturbations}
The relative density perturbation $\delta\rho$ 
and the non-axisymmetric density fluctuation $\Delta\rho$ are defined as 
\begin{align}
  \delta\rho(\bm{r},t) \equiv \frac{\rho - \rho_i}{\rho_i}, \quad
  \Delta\rho(\bm{r},t) \equiv \frac{\rho -
    \aziavg{\rho}}{\aziavg{\rho}}, 
\end{align} 
where $\avg{\cdot}_\phi$ denotes an azimuthal average.  
In general $\Delta\rho$ accounts for the time evolution of 
the axisymmetric part of the density field, but if
$\p_t\avg{\rho}_\phi=0$ then $\Delta\rho$ is identical to
$\delta\rho-\avg{\delta\rho}_\phi$.    

\subsubsection{Vortical structures}
The Rossby number
\begin{align}
  Ro \equiv
  \frac{\hat{\bm{z}}\cdot\nabla\times\bv-\avg{\hat{\bm{z}}\cdot\nabla\times\bv}_\phi}{2\aziavg{\Omega}},   
\end{align} 
can be used to quantify the strength of vortical structures and to
visualize it. $Ro<0$ signifies anti-cyclonic motion with respect to
the background rotation. Note that while for thin discs 
the rotation profile is Keplerian, the shear is non-Keplerian for radially  
structured discs (i.e. $\Omega\simeq\Omega_k$ but the epicycle
frequency $\kappa\neq\Omega$). 

\subsubsection{Potential vorticity}
The potential vorticity (PV, or vortensity) is 
$  \bm{\eta}_\mathrm{3D} = \nabla\times~\bm{v}/\rho$. 
However, it will be convenient to work with vertically averaged
quantities. We define  
\begin{align}
  \eta_z = \frac{1}{\Sigma}\int \hat{\bm{z}}\cdot\nabla\times\bv dz
\end{align}
as the PV in this paper, where $\Sigma = \int\rho dz$ and the
integrals are confined to the computational domain. 
We recall for a 2D disc the vortensity is defined as 
$\eta_\mathrm{2D} \equiv \hat{\bm{z}}\cdot\nabla\times\bv/\Sigma$, and
extrema in $\eta_\mathrm{2D}$ is necessary for the RWI in
2D \citep{lovelace99,lin10}. If the velocity field is independent of 
$z$ then $\eta_z$ is proportional to $\eta_\mathrm{2D}$ (at fixed
cylindrical radius).    

\subsubsection{Perturbed kinetic energy density}  
We define the perturbed kinetic energy as
$W\equiv\rho|\bm{v}|^2/[\rho_i|\bm{v}(t~=0)|^2] - 1$, and its Fourier transform 
$W_m\equiv\int_0^{2\phi} W\exp{(-\mathrm{i}m\phi)}d\phi$. We will examine
$|W_m(r,\theta)|$ averaged over sub-portions of the $(r,\theta)$
plane. 

\section{Non-axisymmetric instability of artificial {\bf radial} density bumps in
  layered discs}\label{density_bump} 
We first consider strictly isothermal discs ($q=0$) initialised with a
density bump ($A>1$). Our aim here
is to examine the effect of (layered) viscosity on the RWI 
\emph{through the linear perturbation}. 
In general a density bump in a viscous disc will undergo viscous
spreading \citep{lyndenbell74}, but we can circumvent this 
by choosing the viscosity profile $\nu$ and initial cylindrical radial
velocity $v_{Ri}$ appropriately. 
Although artificial, this setup avoids 
the simultaneous evolution of the density bump subject to
axisymmetric viscous spreading and growth of
non-axisymmetric disturbances; only the latter of which is our focus.

\subsection{Viscous equilibria for a radially structured
  disc}\label{visc_eq}  
In choosing $\rho_i$ and $\Omega_i$, we neglected radial
velocities and viscous forces in the steady-state vertical and
cylindrical radial momentum equations (Eq. \ref{vert_balance} and
Eq. \ref{init_vphi}, respectively). This is standard practice for
accretion disc modeling \citep[e.g.][]{takeuchi02}.   

However, $v_{R}$ and $\nu$ cannot be ignored in the azimuthal
momentum equation. Indeed, if a steady-state is desired, then the
conservation of angular momentum in a viscous disc implies special
relations between the viscosity, cylindrical radial velocity and
density field. 

\subsubsection{Initial cylindrical radial velocity}
For axisymmetric flow with $\Omega=\Omega(R)$, the azimuthal momentum
equation reads  
\begin{align}\label{ang_mom_cons}
  R\rho v_R\frac{\p}{\p R}\left(R^2\Omega \right) = \frac{\p}{\p
    R}\left(R^3\rho\nu\frac{\p\Omega}{\p R}\right). 
\end{align}
Note that the viscous term due to vertical shear ($\partial_z\Omega$) 
is absent because in this experiment we are considering barotropic
discs. Assuming a steady state with $v_z=0$, mass
conservation (Eq. \ref{cont_eq}) implies that the mass flux  
$\dot{M}\equiv R\rho v_R$ is independent of $R$. In this case,
Eq. \ref{ang_mom_cons} can 
be integrated once to yield 
\begin{align}\label{temp}
  \dot{M}R^2\Omega = R^3\rho\nu\Omega^\prime + C(z) \quad \text{if $\p_R\dot{M}$} = 0, 
\end{align}
where $^\prime$ denotes $d/dR$ and $C(z)$ is an arbitrary function of
$z$. Eq. \ref{temp} motivates the simple choice
\begin{align}\label{init_vr} 
  v_{Ri} = \frac{\nu}{R}\frac{d\ln{\Omega_i}}{d\ln{R}} 
\end{align}
for the initial cylindrical radial velocity. Next, we choose the
viscosity profile $\nu$ to make $\dot{M}$ independent of $R$.   

\subsubsection{Viscosity profile for a steady state}\label{visc_model}
If the initial conditions corresponds to a steady state, then
$R \rho_i v_{Ri}$ can only be a function of $z$. With $v_{Ri}$ chosen
by Eq. \ref{init_vr}, this implies 
$R\rho_i\nu\Omega_i^\prime/\Omega_i$ is only a function of $z$. We are
therefore free to choose the vertical dependence of viscosity.   

Let $\nu = \hat{\nu}r_0^2\Omega_k(r_0)$, where
$\hat{\nu}=\hat{\nu}(R,z)$ is a dimensionless function describing
the magnitude and spatial distribution of the axisymmetric kinematic
viscosity. We choose $\hat{\nu}$ such that   
\begin{align}\label{visc_profile}
  \hat{\nu}\rho_i(R,z)\frac{d\ln{\Omega_i}}{d\ln{R}} =
  \hat{\nu}_0\left[1+Q(z/H_0)\right]\rho_i(r_0,z)\left.\frac{d\ln{\Omega_i}}{d\ln{R}}\right|_{r_0}, 
\end{align}
where $\nu_0$ is a constant dimensionless floor viscosity and   
\begin{align}\label{step}
  Q(\zeta) = \frac{\left(A_\nu - 1\right)}{2}
  \left[  2 + \tanh{\left(\frac{\zeta - \zeta_\nu}{\Delta\zeta_\nu}\right)}
    - \tanh{\left(\frac{\zeta +
        \zeta_\nu}{\Delta\zeta_\nu}\right)}\right]
\end{align}
is a generic function describing a step of magnitude
$A_\nu-1$. The position and width of the step is described by
$\zeta_\nu$ and $\Delta\zeta_\nu$, respectively, with $\Delta\zeta_\nu\ll\zeta_\nu$. 
In Eq. \ref{visc_profile} we have set the dimensionless co-ordinate
$\zeta=z/H_0$ where $H_0=H(r_0)$. 
We can translate $\hat{\nu}$ to an alpha viscosity 
using $\nu = \alpha c_s H$ \citep{shakura73} so that $\alpha =
\hat{\nu}/h^2$ at $R=r_0$.  This gives $\alpha\sim 10^{-2}$ for
$h=0.1$ and $\hat{\nu}=10^{-4}$. 

\begin{figure}
  \centering
  \includegraphics[width=\linewidth]{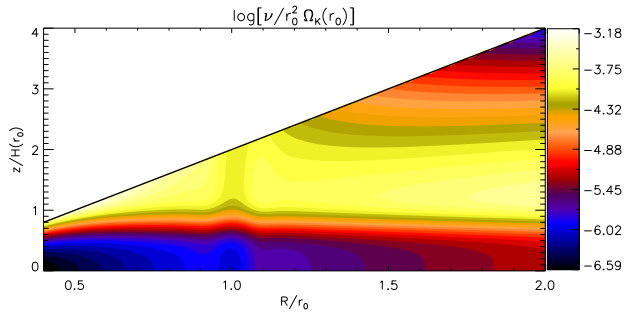}
  \caption{Example of a two-layered kinematic viscosity profile
    resulting from Eq. \ref{visc_profile}. This specific plot
    corresponds to case V2. The solid line
    delineates the upper boundary of the computational domain.
    \label{visc2d}}
\end{figure}

Eq. \ref{visc_profile} implies that at the fixed cylindrical radius
$R=r_0$, the dimensionless viscosity increases from $\hat{\nu} =
\hat{\nu}_0$ at the midplane to $\hat{\nu} = A_\nu\hat{\nu}_0$ for
$z > \zeta_\nu H_0$. An example of such a layered viscosity profile
profile is depicted in Fig. \ref{visc2d}.

\subsection{Simulations}
We consider discs with radial extent $[\rin,\rout]=[0.4,2.0]r_0$,
vertical extent $n_h=2$ scale-heights and aspect-ratio $h=0.1$ at
$R=r_0$. We use $(N_r, N_\theta,
N_\phi)=(256,64,512)$ grid points. 
The resolution at the reference radius is then
$16,\,32,\,8$ cells per scale-height in $(r,\theta,\phi)$ directions,
respectively. The planet potential is disabled for these runs
($M_p\equiv 0$). We apply a damping rate $\hat{\gamma}=1$ with the
reference velocity $\bm{v}_\mathrm{ref}=\bm{v}(t=0)$.     

\begin{figure}
  \centering
  \includegraphics[width=\linewidth]{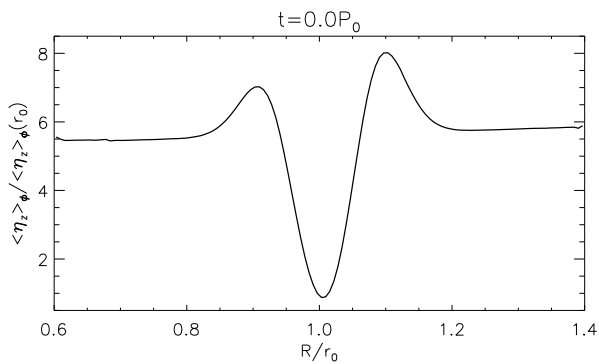}
  \caption{  Potential vorticity profile for simulations initialised
    with a surface density bump, as described by
    Eq. \ref{bump_func}. The RWI is associated with the local PV
    minima at unit radius.
    \label{bump_PV}}
\end{figure}

The bump parameters are set to $A=1.25$ and $\Delta R = 0.05r_0$ for
all runs in this section.  The corresponding PV profile is shown
  in Fig. \ref{bump_PV}. The spherical radial velocity is subject 
to random perturbations of magnitude $10^{-4}c_s$ 
a few time-steps after initialization.

\subsubsection{Linear growth rates and frequencies}
The present setup allows us to define a linear instability in the
usual way: exponential growth of perturbations measured with respect
to an axisymmetric steady equilibrium. A proper linear
stability analysis, including the full viscous stress tensor, is
beyond the scope of this paper, but we can nevertheless extract linear 
mode frequencies from the non-linear simulations.  

The $m^\mathrm{th}$ Fourier component of the density field is  
\begin{align}
\hat{\rho}_m(r,\theta,t) \equiv \int_0^{2\pi} \rho(\bm{r},t)\exp{(-\ii
  m\phi)}d\phi. 
\end{align} 
The magnitude of a Fourier mode is measured by 
\begin{align} 
a_m(t)\equiv \frac{b_m(t)}{b_0(0)},\quad b_m(t)\equiv \avg{|\hat{\rho}_m|}_r,
\end{align} 
where $\avg{\cdot}_r$ denotes averaging over a spherical shell (to be
chosen later). 
The complex frequency $\sigma_m$ associated with the $m^\mathrm{th}$ component 
is defined through 
\begin{align}\label{sigma}
  \frac{\p\hat{\rho}_m}{\p t} \equiv -\ii\sigma_m\hat{\rho}_m. 
\end{align}
The time derivative in Eq. \ref{sigma} can be computed implicitly by 
Fourier-transforming the continuity equation \citep[as done
  in][]{lin13}. 

In a linear stability problem $\sigma_m$ is a constant eigenvalue. 
However, when extracted numerically from a non-linear simulation, 
we will generally obtain $\sigma_m=\sigma_m(r,\theta,t)$. Thus, we compute 
$\avg{\sigma_m}_r = m\omega_m + \ii q_m$    
where $\omega_m$ is the mode frequency and $q_m$ is the growth rate.
We normalize the linear frequencies by
$\Omega_0\equiv\Omega_i(r_0)\simeq\Omega_k(r_0)$. 

\subsection{Results}
Table \ref{artificial_bump} summarizes the simulations presented in
this section. 
For reference we simulate an effectively inviscid disc, case B0,
with the viscosity parameters $\hat{\nu}_0=10^{-9}$ and $A_\nu = 1$.  
Thus viscosity is independent of $z$ at $R=r_0$.   
Inviscid setups similar to case B0 have previously been simulated
both in the linear and non-linear regimes  \citep{meheut12, lin13}. 

We then simulate discs with floor viscosity  
$\hat{\nu}_0=10^{-6}$. The control run case V0 has $A_\nu =
1$. Thus, case V0 is the viscous version of case B0.  
We then consider models where the kinematic viscosity increases by
a factor $A_\nu=100$ for $z>\zeta_\nu H_0$ at the bump radius. We
choose $\zeta_\nu=1.5,\,1.0$ for cases V1 and V2, respectively.  This
gives a upper 
viscous layer of thickness $0.5H$ and $H$ at $R=r_0$. (See
Fig. \ref{visc2d} for a plot of the kinematic viscosity profile for case V2.) 
For case V1 and V2 the transition thickness is fixed to
$\Delta\zeta_\nu=0.2$. Finally, we consider a high viscosity run, case
V3, with $\hat{\nu}_0=10^{-4}$ and $A_\nu=1$.  This is equivalent to 
extending the viscous layer in case V1/V2 to the entire vertical
domain.  

\begin{table*}
  \centering
  \caption{Summary of hydrodynamic simulations initialized with a
    density bump. Linear mode frequencies and the non-linear mode amplitudes $a_m$ are averaged over 
    $r\in[0.8,1.2]r_0$. \label{artificial_bump}}
  \begin{tabular}{lcccccl @{\extracolsep{0.1cm}} ccc}
    \hline\hline
    \multicolumn{4}{c}{\phantom{stuff}} &
    \multicolumn{3}{c}{$t = 10P_0$ (linear phase)}&
    \multicolumn{3}{c}{$t=100P_0$}\\
    \cline{5-7}\cline{8-10}
    Case  & $\log{\hat{\nu}_0}$ & $A_\nu$ &$\zeta_\nu$ & $m$ &
    $\omega_m/\Omega_0$ &
    $q_m/\Omega_0$ &  
    $m$ & $10^2a_m$ & $\mathrm{min}[Ro(z=0)]$ \\ 
    \hline
    B0 &-9 & 1 &n/a & 4 & 0.985  & 0.199  
    &  1 & 8.5  & -0.15   \\  
    
    V0  &-6 & 1 &n/a &  4 & 0.985  & 0.199   
    & 1 & 6.8 &  -0.11  \\
    
    V1  &-6 & 100 & 1.5  & 4 & 0.986  & 0.191
    &  1 & 7.8 &  -0.19 \\
    
    V2  & -6 & 100 & 1.0  &  4  & 0.986  & 0.182  
    &  1 & 4.9 &  -0.21 \\
    
    V3  & -4 & 1 & n/a  &  4  & 0.986  &  0.131  
    &  3 &  3.7  &  -0.25 \\
   \hline
  \end{tabular}
\end{table*}

\subsubsection{Inviscid reference case}
\begin{figure}
  \centering
  \includegraphics[scale=.27,clip=true,trim=0cm 0.9cm 0cm
    0cm]{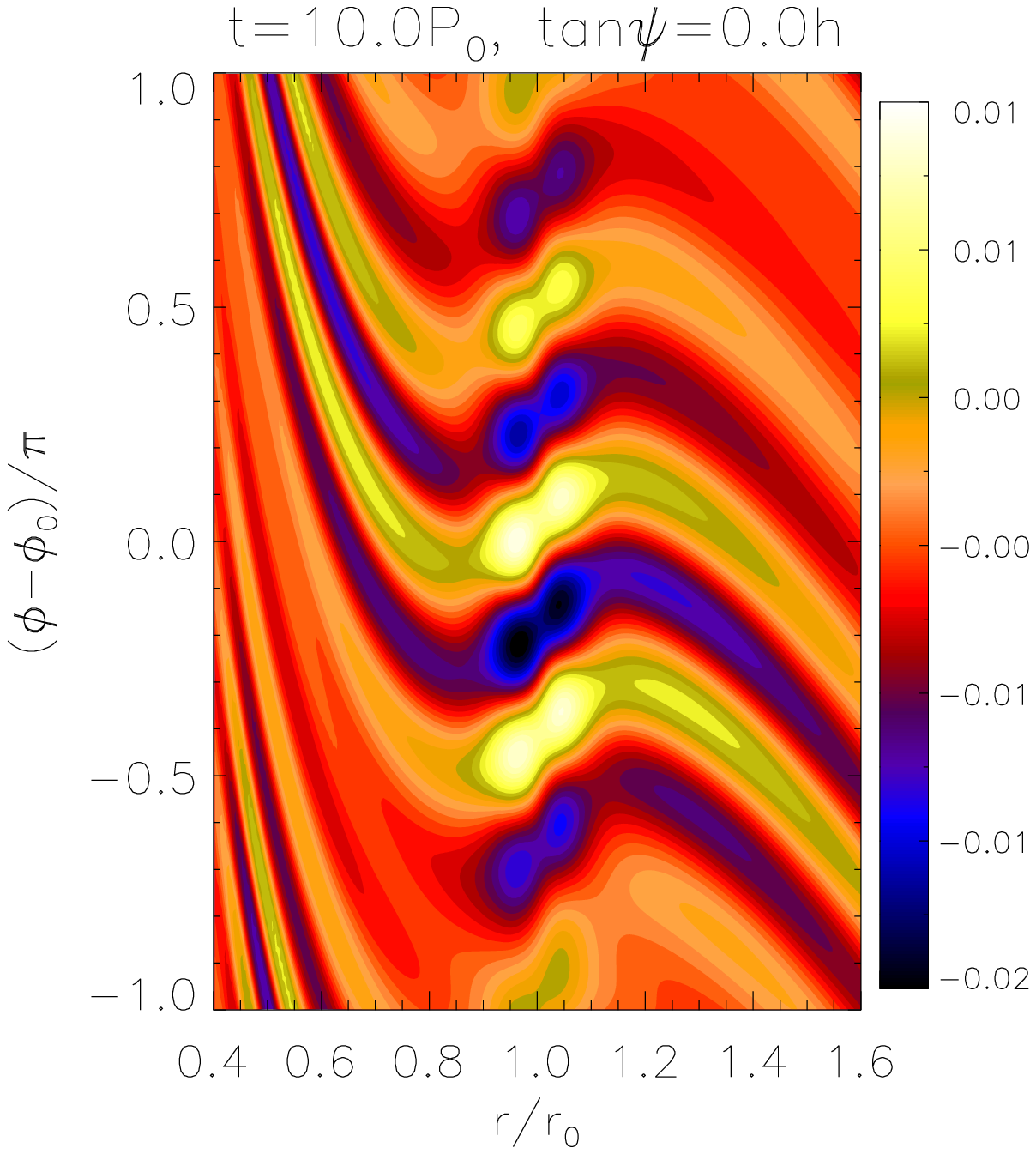}\includegraphics[scale=.27,clip=true,trim=2.3cm
    0.9cm 0cm
    0cm]{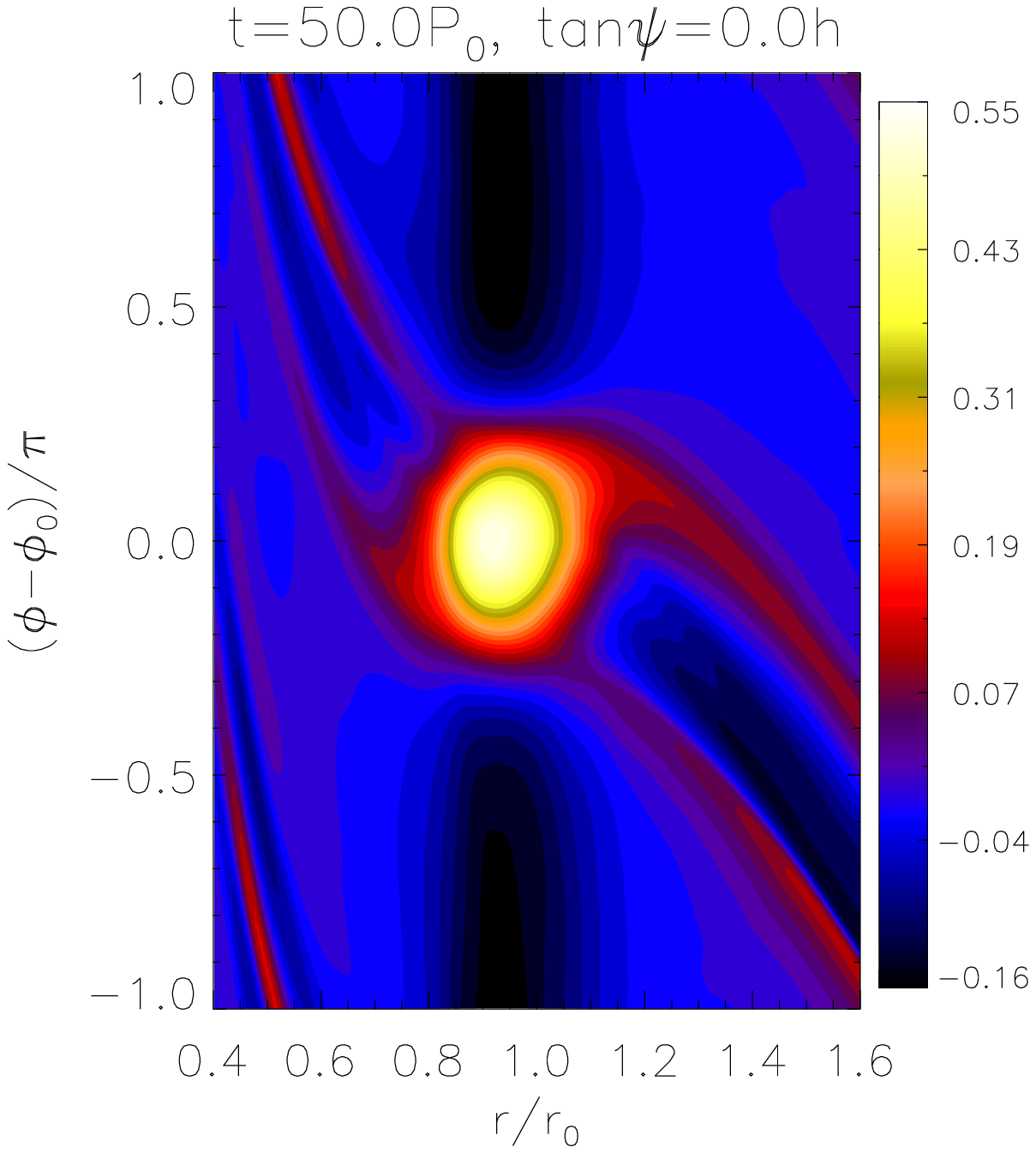}\includegraphics[scale=.27,clip=true,clip=true,trim=2.3cm
    0.9cm 0cm
    0cm]{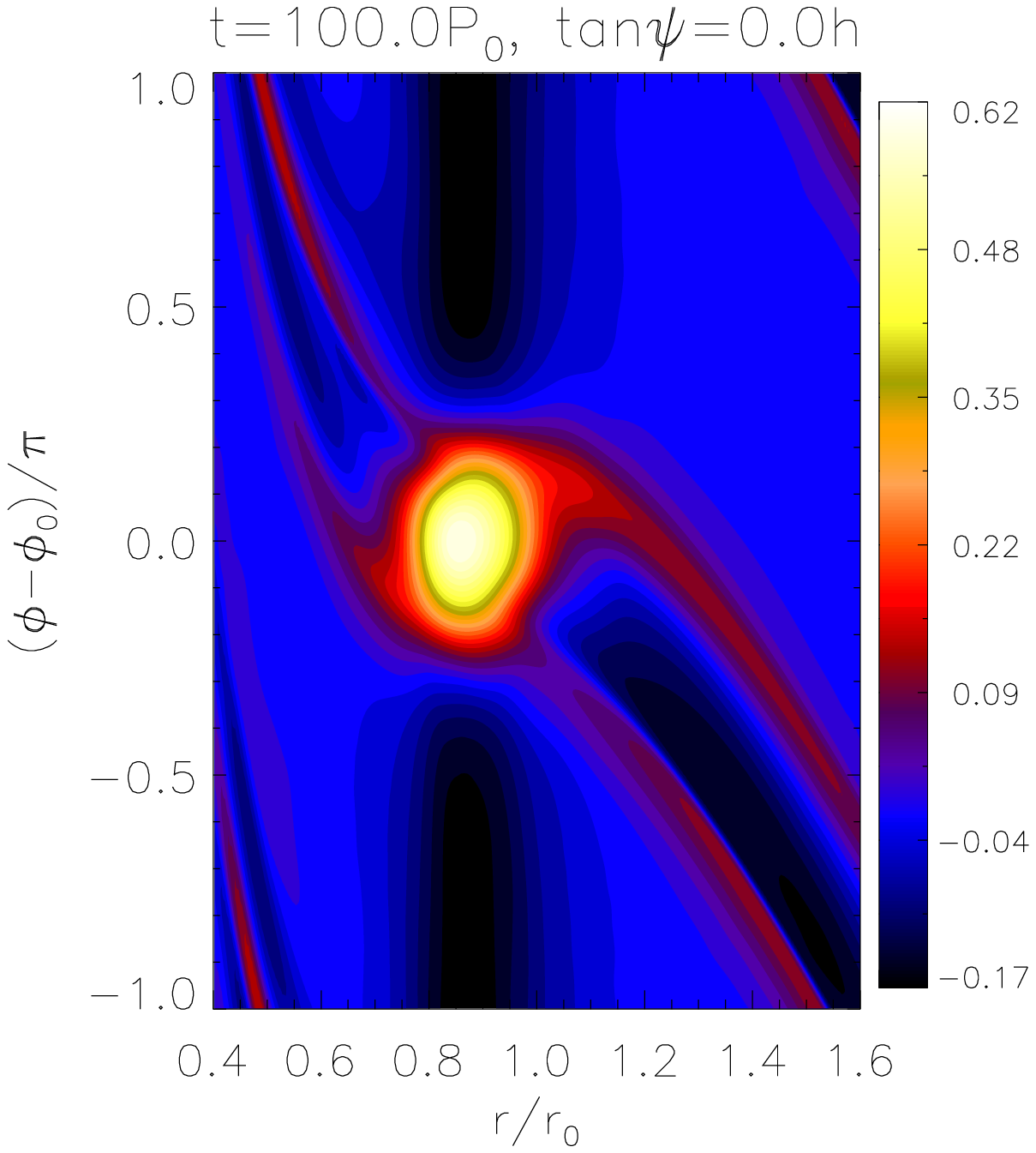}
   \includegraphics[scale=.27,clip=true,trim=0cm 0.cm 0cm
     0.9cm]{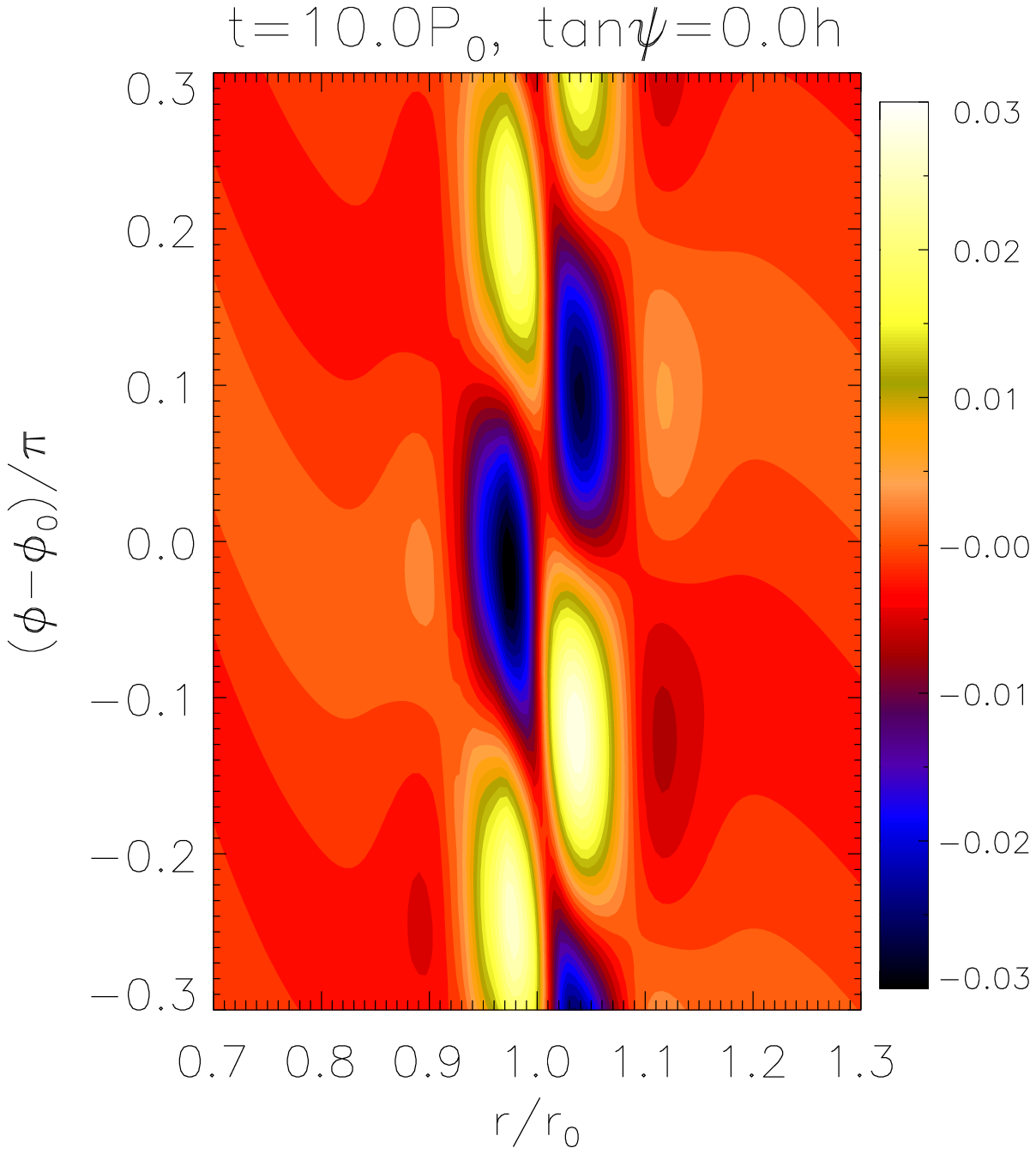}\includegraphics[scale=.27,clip=true,trim=2.3cm
     0.cm 0cm
     0.9cm]{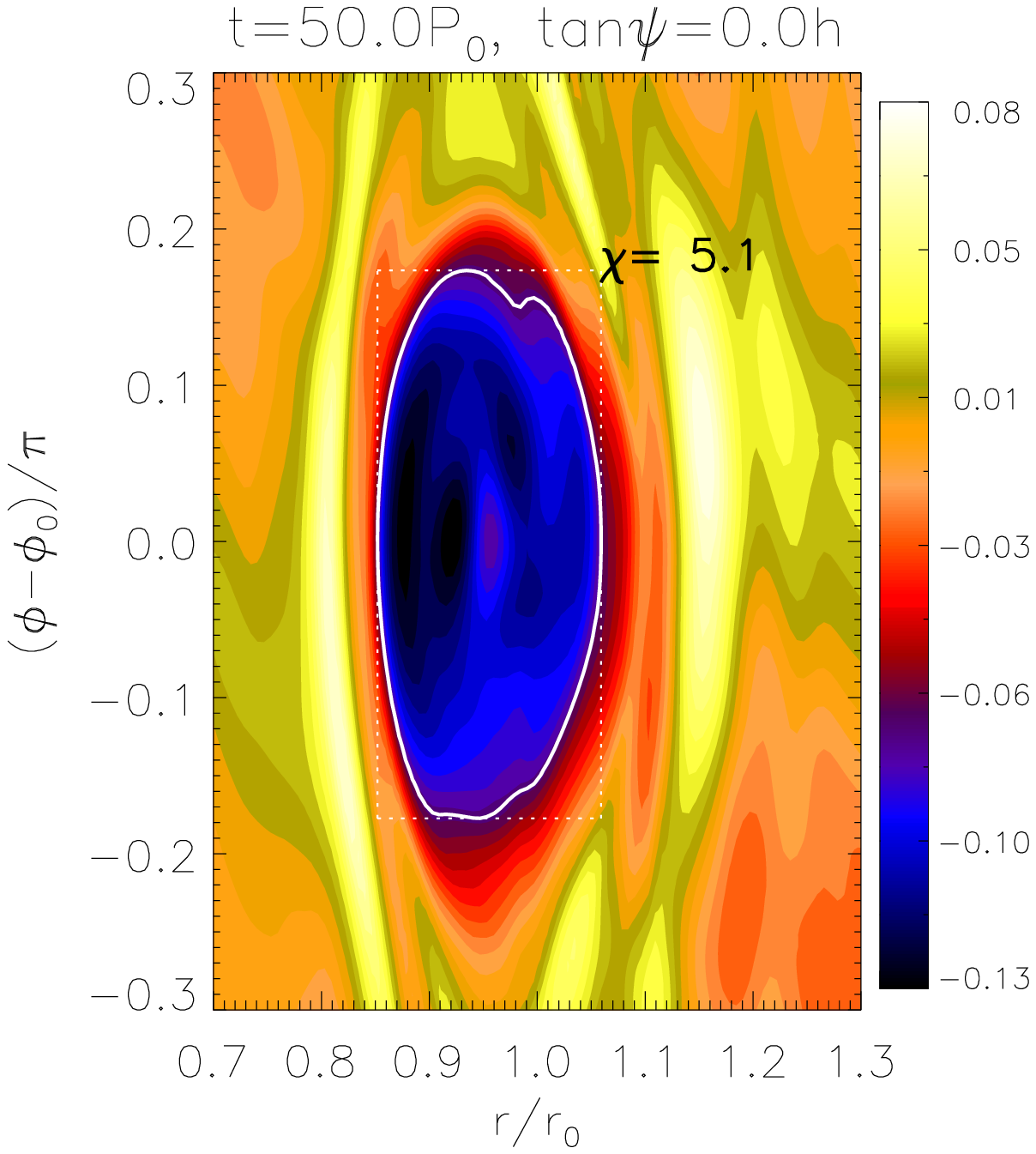}\includegraphics[scale=.27,clip=true,clip=true,trim=2.3cm
     0.cm 0cm
     0.9cm]{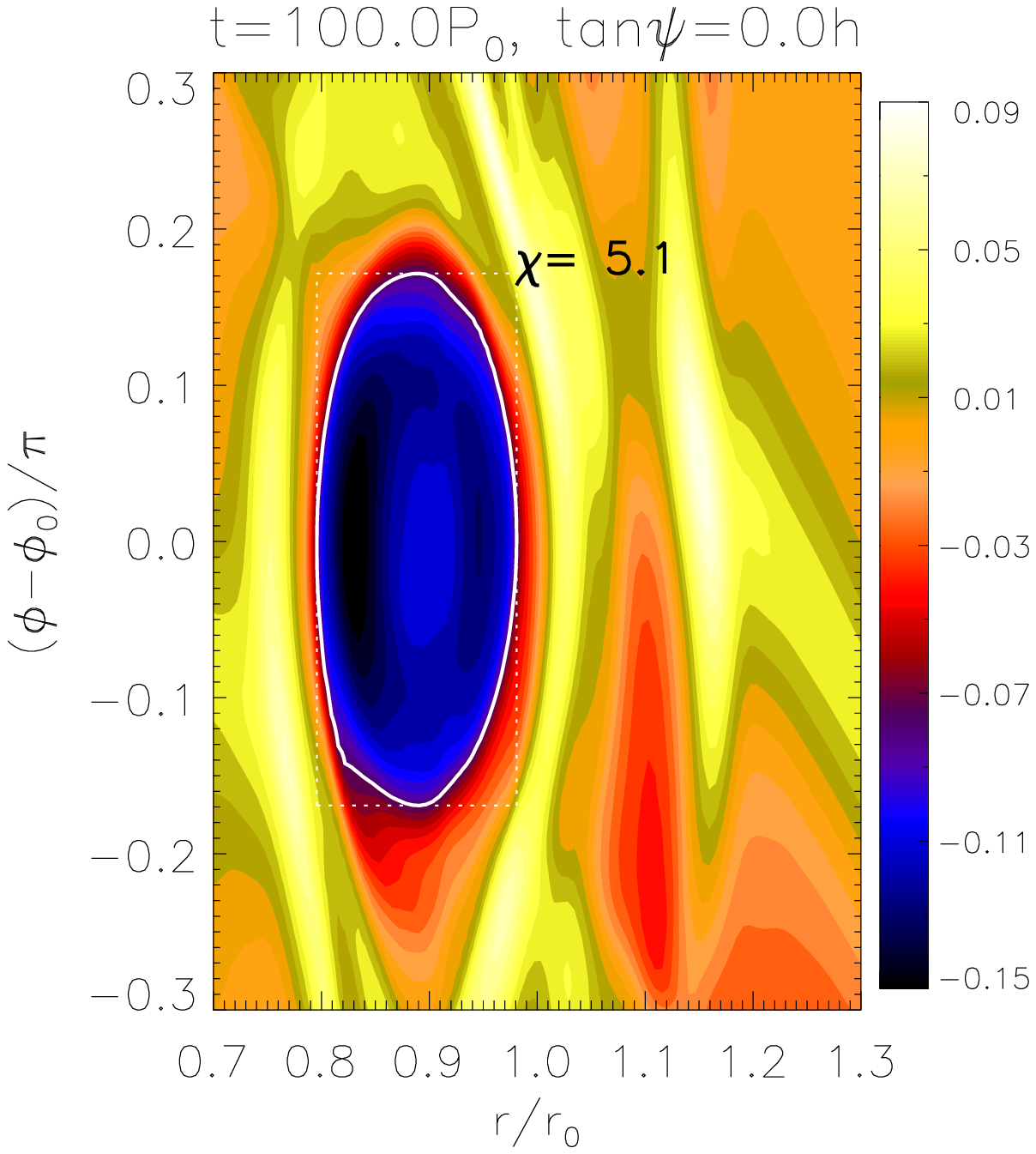}
  \caption{Evolution of the inviscid case B0. Top: midplane density fluctuation, 
    $\Delta\rho(z=0)$. Bottom: midplane
    Rossby number (note the different axis range from the top panel). 
    Here, $\chi$ is an empirical measure of the final vortex
    aspect-ratio. $\phi_0$ is the azimuth of 
    $\mathrm{max}\left[|\Delta\rho(z=0)|\right]$.
    \label{bump0_bump1}}
\end{figure}

Fig. \ref{bump0_bump1} 
shows the density fluctuation and Rossby number for
case B0. The dominant linear mode is $m=4$ with a growth rate
$0.2\Omega_0$, consistent with recent 3D linear calculations 
\citep{meheut12,lin13}. 
The non-linear outcome of the RWI is vortex formation
\citep{li00}. Four vortices develop initially, then 
merge on a dynamical timescale into  
a single vortex. Case B0 evolves similarly to previous simulations of
the RWI in an inviscid disc 
\citep[e.g.][where more detailed analyses are 
given]{meheut10,meheut12b}. This, together with the agreement with
linear calculations, demonstrates the ability of the \pluto
code to capture the RWI. 

\subsubsection{The effect of a viscous layer}
We now examine viscous cases V0 --- V3. Recall  
from Table \ref{artificial_bump} that the viscous layer (with $
\hat{\nu}\sim10^{-4}$) occupies the uppermost $0\%,\,25\%,\,50\%$ and
$100\%$ of the vertical domain at $R=r_0$ for cases V0, V1, V2 and V3,
respectively.     

\begin{figure*}
   \centering
   \includegraphics[scale=.43,clip=true,trim=0cm 0.9cm 0cm
     0.99cm]{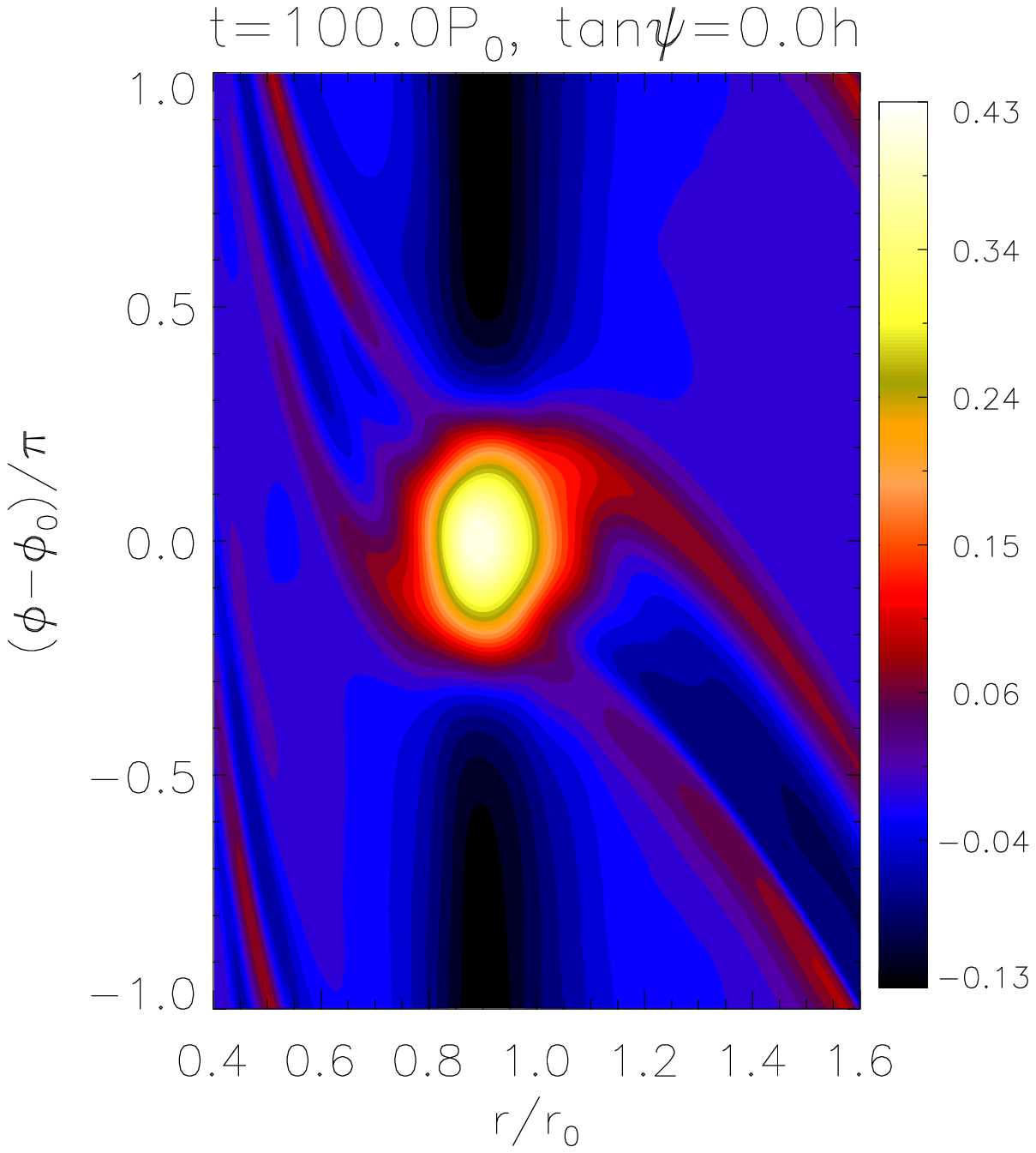}\includegraphics[scale=.43,clip=true,trim=2.3cm
     0.9cm 0cm
     0.99cm]{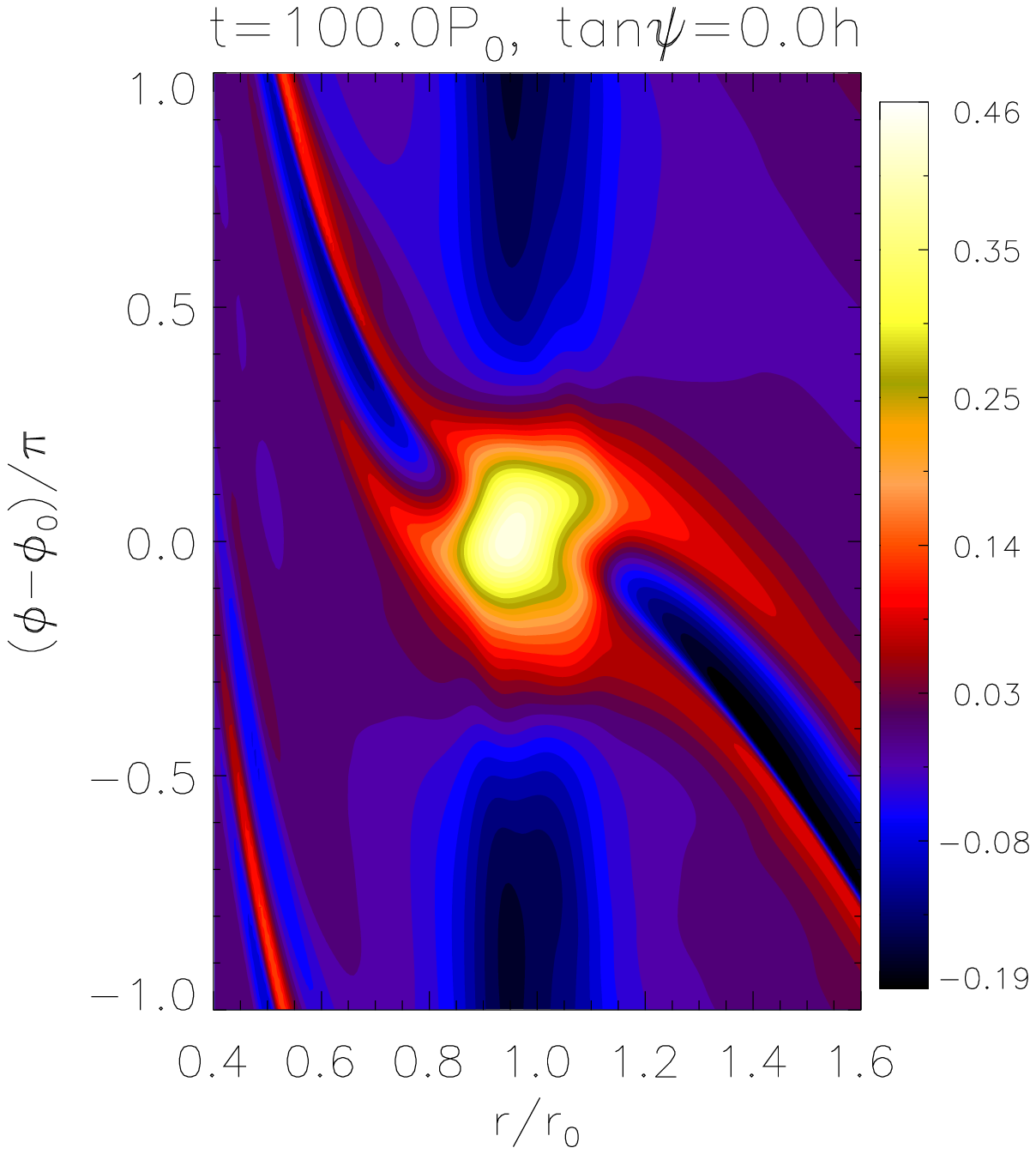}\includegraphics[scale=.43,clip=true,trim=2.3cm 
     0.9cm 0cm
     0.99cm]{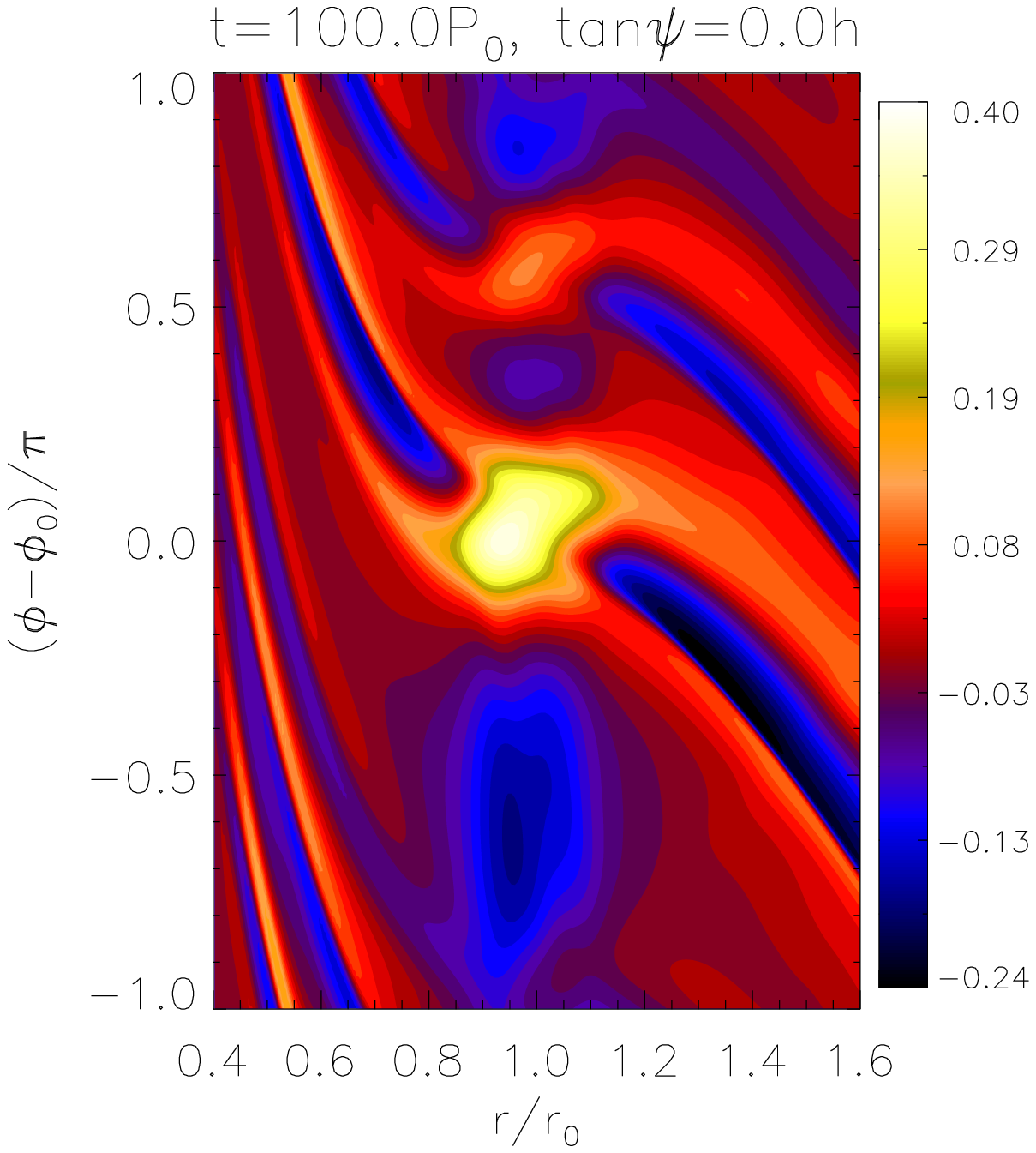}\includegraphics[scale=.43,clip=true,trim=2.3cm
     0.9cm 0cm
     0.99cm]{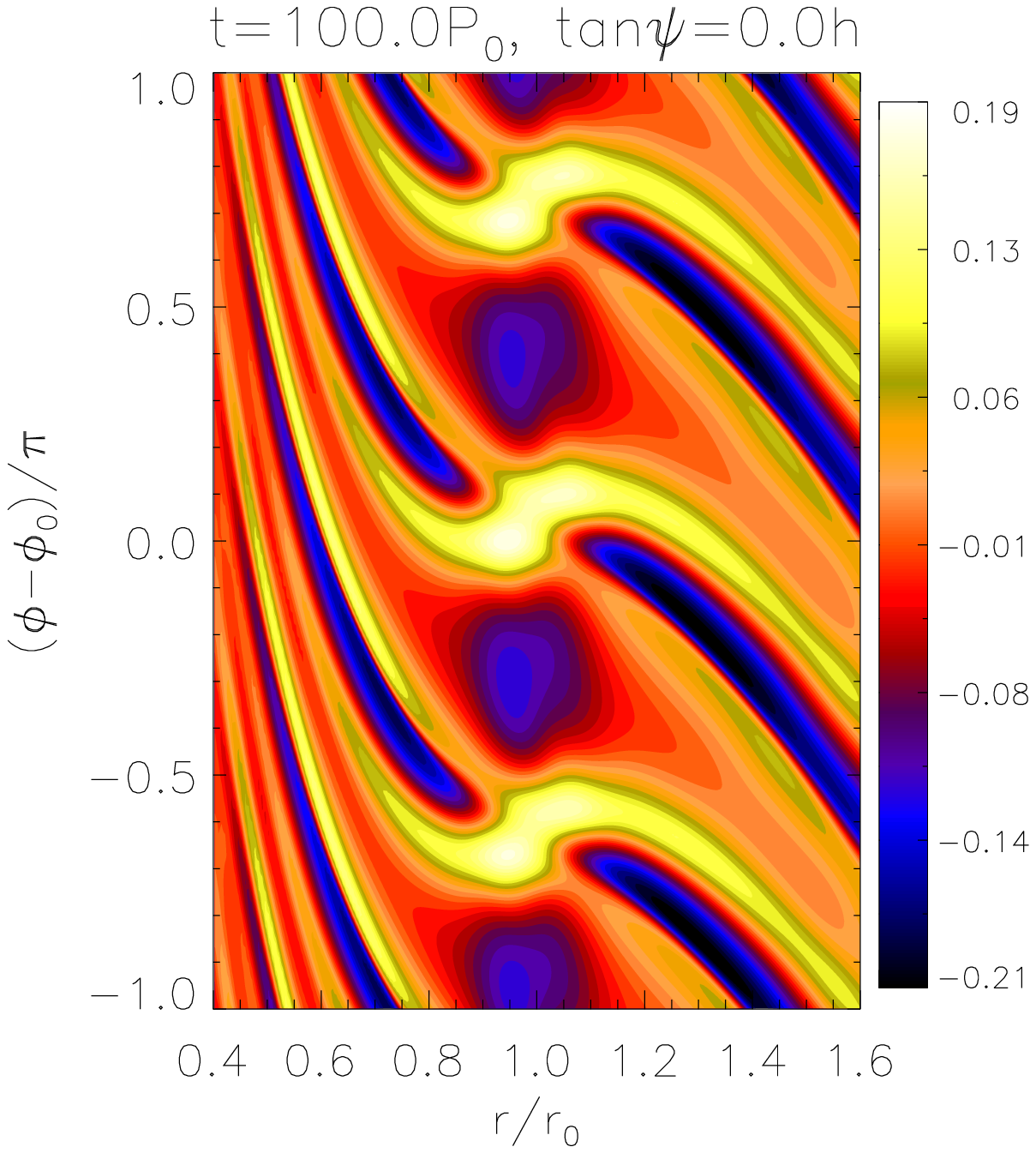}\\
      \includegraphics[scale=.43,clip=true,trim=0cm 0.cm 0cm
     0.99cm]{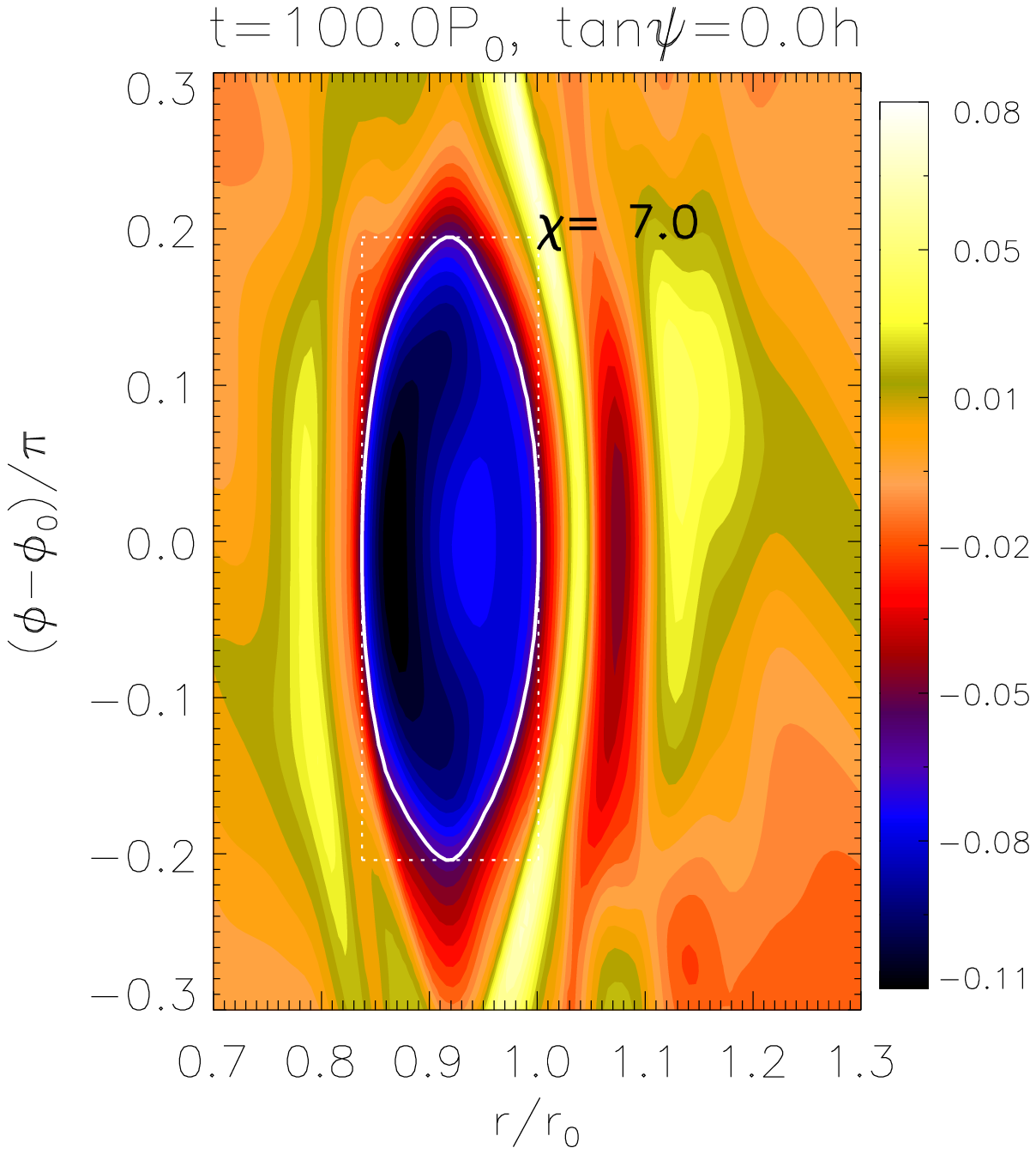}\includegraphics[scale=.43,clip=true,trim=2.3cm
     0.0cm 0cm
     0.99cm]{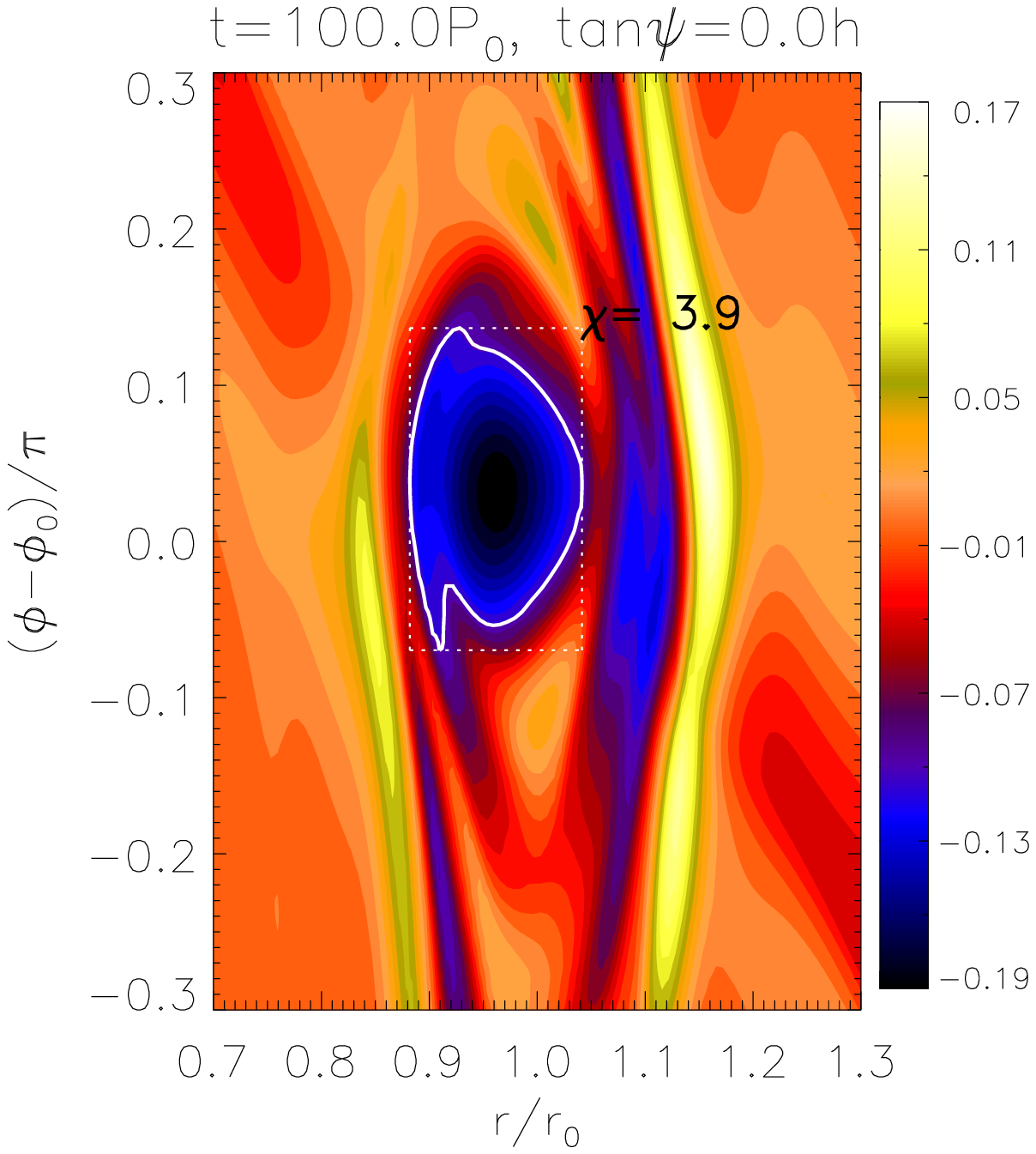}\includegraphics[scale=.43,clip=true,trim=2.3cm 
     0.0cm 0cm
     0.99cm]{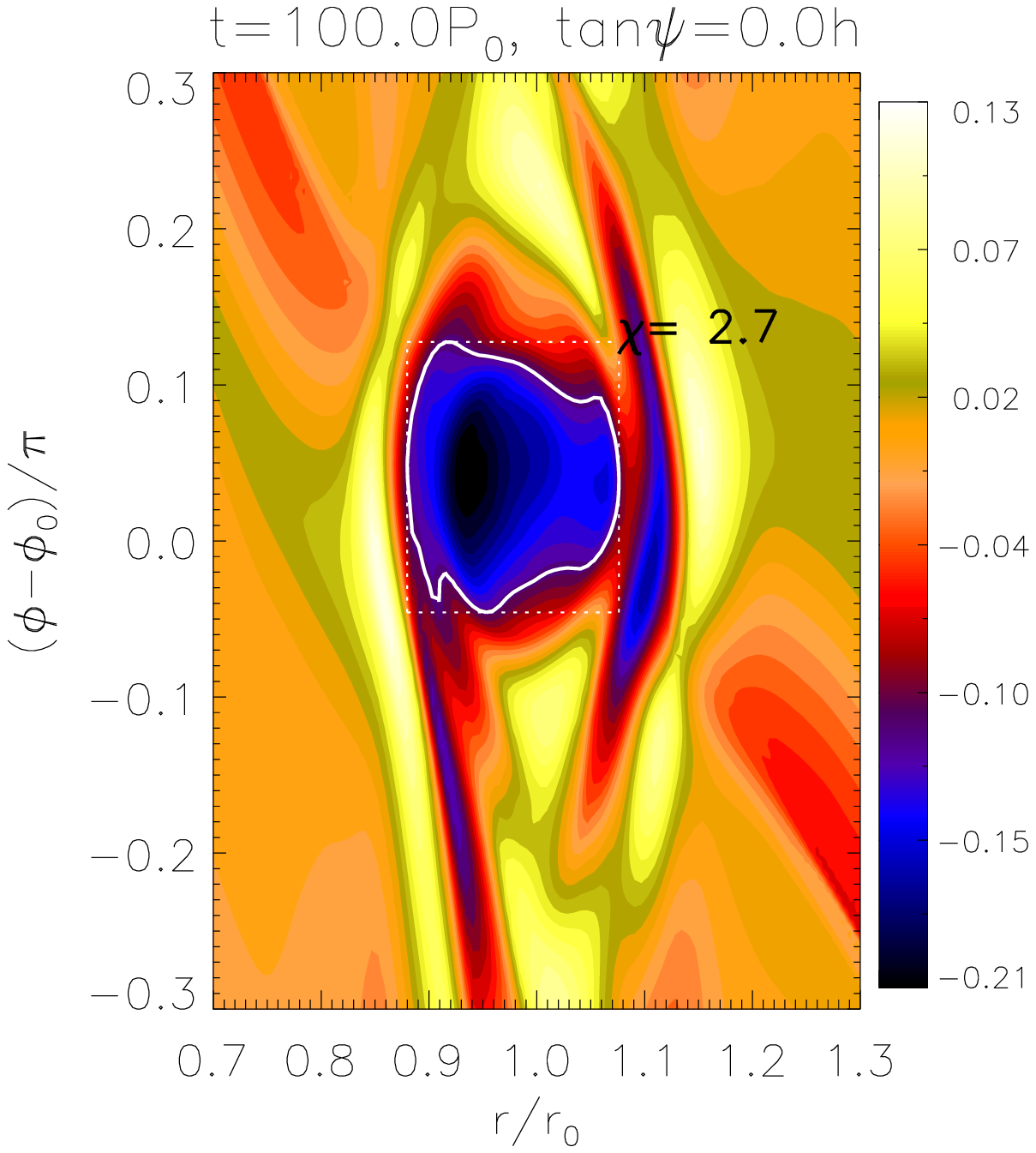}\includegraphics[scale=.43,clip=true,trim=2.3cm
     0.0cm 0cm
     0.99cm]{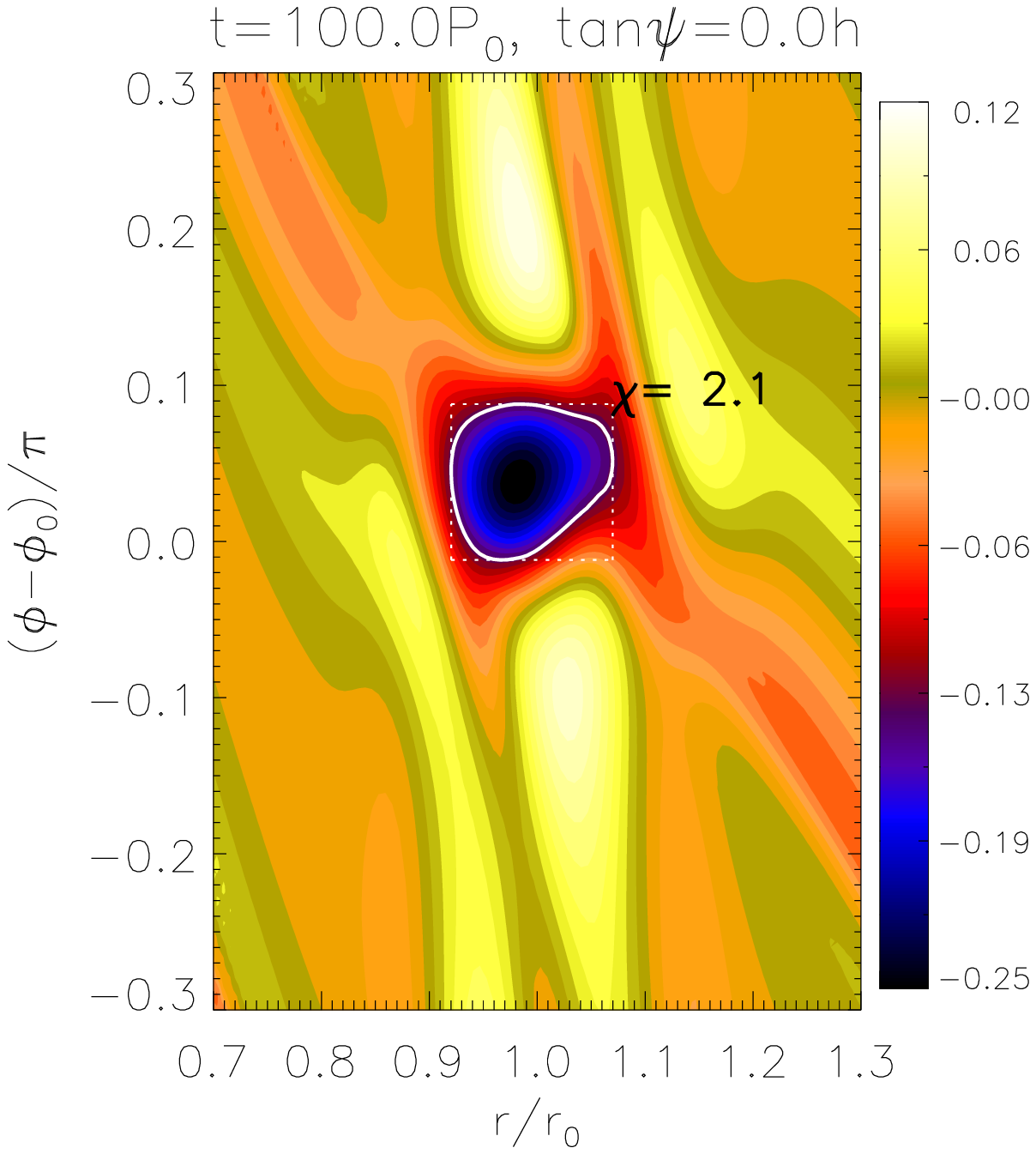}
   \caption{Vortex formation in viscous discs initialised with
     a density bump at unit radius. Snapshots are taken at $t=100P_0$. The
     thickness of the viscous layer increases from left to 
     right: case V0, V1, V2 and V3.   
     Top: non-axisymmetric density field at the midplane
     $\Delta\rho(z=0)$. Bottom: midplane Rossby number
     $Ro(z=0)$. Here, $\phi_0$ is the azimuth of $\max[\Delta\rho(z=0)]$. 
     \label{vdamp0}
   }
\end{figure*}

We first compare the viscous case V0 to the 
inviscid case B0. Table \ref{artificial_bump} shows that despite
increasing the viscosity by a factor of $10^3$, the change to the
linear mode frequencies are negligible. 
The value of $a_m$ and minimum Rossby number indicate that the final
vortex in V0 is only slightly weaker than that in B0. This is also
reflected in  Fig. \ref{bump0_bump1} (case B0) and the leftmost column in
Fig. \ref{vdamp0} (case V0). Case V0 develops a more elongated 
vortex with smaller $|\Delta\rho|$ than that in B0. 
  
As we introduce and thicken the viscous layer from case V0 to V3, the dominant
linear mode remains at $m=4$ (Table \ref{artificial_bump}), but linear
growth rate does appreciably decrease (by $\sim 34\%$ from case V0 to
V3). However, these linear growth timescales  
are still $\sim P_0$. We thus have the important result that
viscosity (layered or not) does not significantly affect the linear
instability because the RWI grows dynamically even in the high
viscosity disc.     
 
The effect of layered viscosity in the non-linear regime is more
complicated. The bottom row of Fig. \ref{vdamp0} compares the Rossby number
associated with the over-densities. Thickening the viscous 
layer decreases the vortex aspect ratio. Since their widths remain
at $\sim 2H_0$, the vortices become smaller with increasing
viscosity. This is partly attributed to fewer vortex merging
events having occurred as viscosity is increased, which 
usually results in larger but weaker vortices (smaller
$|Ro|$). If merging is resisted then each vortex can grow individually.  
Strangely, vortices become stronger  (more negative $Ro$) as viscosity
is increased. 

\begin{figure}
  \centering
  \includegraphics[width=\linewidth]{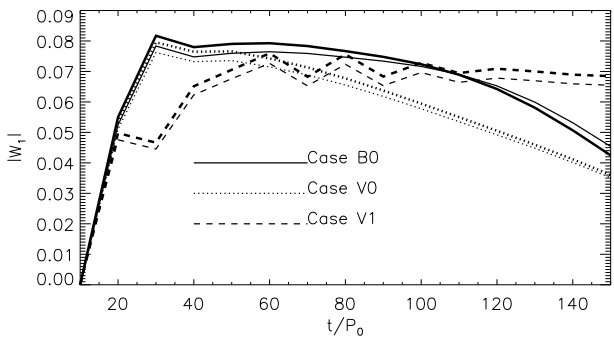}
  \caption{The $m=1$ component of the kinetic energy density, averaged
    over $r\in[0.8,1.2]r_0$, for the inviscid case B0 (solid), low
    viscosity case V0 (dotted) and a layered viscosity case V1 (dashed). 
    For each, the contribution 
    averaged over the disc atmosphere ($\tan{\psi}\in[1.5,2.0]h$, thin
    lines) and over the
    disc bulk ($\tan{\psi}\in[0,1.5]h$, thick lines) are plotted
    separately. 
    \label{bump_energy}}
\end{figure}

Fig. \ref{bump_energy} compares the perturbed kinetic energy for 
cases B0, V0 and V1; which are all dominated by a single vortex in 
quasi-steady state. We compute $W_1$ and compare its
average over the disc atmosphere and over the 
disc bulk. There is only a minor difference between the
perturbed kinetic energy densities between the disc bulk and
atmosphere, even in case V1 where the kinematic viscosity in the two
regions differ by a factor $\sim10^2$. This suggests that the
vortex evolves two-dimensionally. 

The energy perturbation in case B0 and V0 are both subject to
slow decay \citep[a result also observed by][]{meheut12}.  
By contrast case V1, which includes a high viscosity layer, does
\emph{not} show such a decay. We discuss this counter-intuitive result
below.  

\subsection{Order of magnitude comparison of timescales}

The characteristic spatial scale of the background density bump and of
the RWI is the local scale-height $H$, so the associated
viscous timescale is    
\begin{align}
  t_\nu = \frac{H^2}{\nu}\sim \frac{h^2}{\hat{\nu}\Omega}. 
\end{align}
The linear instability growth timescale is
\begin{align}
  t_\mathrm{RWI} = \frac{1}{\epsilon \Omega},
\end{align}
where $\epsilon$ is found from numerical simulations.   
The ratio of these timescales is
\begin{align}
  \frac{t_\nu}{t_\mathrm{RWI}} \sim \frac{\epsilon h^2}{\hat{\nu}}.
\end{align}
Table \ref{artificial_bump} indicates $\epsilon \sim 0.1$. 
Inserting $h=0.1$ and $\hat{\nu}=10^{-4}$ gives $t_\nu \sim 10
t_\mathrm{RWI}$. Thus viscosity damping is slower than linear
growth, even for the highest viscosity values we consider. 
Consequently the linear RWI is unaffected by viscosity. 

\cite{meheut13} argued that $t_\mathrm{RWI}$ is also the vortex
turn-over time $t_\mathrm{turn}$ when the instability saturates and
the linear phase terminates. Then 
$t_\nu\sim 10 t_\mathrm{turn}$, implying viscous effects 
are unimportant over one turn-over time. 
However, if we estimate a vortex turn-over time as $t_\mathrm{turn}
\sim 2\pi/|Ro|\Omega$ then $t_\nu\sim (h^2|Ro|/2\pi\hat{\nu})t_\mathrm{turn}$.  
Inserting $h=0.1,\,\hat{\nu}=10^{-4}$ and $|Ro|=0.25$ (case V3) gives 
$t_\nu \sim 4t_\mathrm{turn}$. Therefore, depending on the vortex shape, 
$t_\nu$ may not be much larger than $t_\mathrm{turn}$. 

In any case, our high-viscosity simulations span several local viscous timescales, 
$t_\mathrm{sim}\sim 10t_\nu $ (for $\hat{\nu}\sim10^{-4}$), so viscous
damping should have taken 
place, making the observation that $Ro$ becomes more negative as the 
viscous layer increases from case V0 to V3, a surprising
result. However, recall that we imposed a stationary, radially
structured viscosity profile consistent with a steady-state disc
containing a density bump. We suggest that for such setups, viscosity
attempts to restore the initial disc profile, i.e. the initial PV
minimum, thereby acting as a vorticity source.      

\subsection{Potential vorticity evolution}
The RWI is stronger for deeper PV minima \citep{li00}. We 
thus expect deeper PV minima to correlate with stronger vortices.   
For the above simulations the axisymmetric PV perturbation at the bump
radius is   
\begin{align}\label{vortensity_pert}
  \left.\frac{\aziavg{\eta_z}(t=100P_0)}{\eta_z(t=0)}\right|_{R=r_0} - 1 = 
  \begin{cases}
    3.99  & \text{Case V0} \\
   3.03  & \text{Case V1} \\
   2.24  & \text{Case V2} \\
   1.71  & \text{Case V3} \\
  \end{cases}.
\end{align} 
(This value is 4.92 for the inviscid case B0.) The PV 
perturbation is positive, thus the initial PV minimum is weakened
by the vortices \citep{meheut10}. This effect
diminishes with increasing viscosity. One contributing factor 
is the reduction in linear growth rate (Table 
\ref{artificial_bump}), implying the  
instability saturates at a smaller amplitude \citep{meheut13}. This  
is expected to weaken the background axisymmetric structure to a lesser
extent. However, the imposed viscosity profile may also actively
restore the initial density bump.     

When viscosity is small, the local viscous timescale $t_\nu$ is long compared to our
simulation timescale $t_\mathrm{sim}$. Then vortex formation weakens the PV minimum
with viscosity playing no role. Increasing viscosity eventually makes 
$t_\nu<t_\mathrm{sim}$. This means that over the course of the
simulation, our spatially-fixed viscosity profile can act to recover the
initial PV minimum. 

 We also notice reduced vortex migration in Fig. \ref{vdamp0}
  with increased viscosity (e.g. the vortex in case V0 has migrated
  inwards to $R\simeq0.9r_0$ while that in case V3 remains near
  $R\simeq r_0$). \cite{paardekooper10} have shown that vortex migration can be halted by
  a surface density bump which, in our case, can 
  be sourced by the radially-structured viscosity profile.   

We conjecture that in the non-linear regime there is competition 
between destruction of the background PV minimum by the 
vortices and reformation of the initial radial PV minimum by the
imposed viscosity profile. The latter effect should favour the RWI,
since the anti-cyclonic vortices are regions of local vorticity
minima. In this way, viscosity acts to source vorticity, and this
effect outweighs viscous damping of the linear perturbations. We
  discuss additional simulations supporting this hypothesis in
  Appendix \ref{add_sim}.

\section{Vortex formation at planetary gap edges in layered
  discs}\label{disc-planet} 
The previous simulations, while necessary to isolate the effect of 
viscosity on the linear RWI, has the disadvantage that the radially
structured viscosity profile can act to source radial disc structure
in the non-linear regime. In this section, we employ a radially smooth
viscosity profile and use 
disc-planet interaction to
create the disc structure required for instability. Then we expect viscosity to only
act as a damping mechanism. 

Vortex formation at gap edges is a standard result in 
2D and 3D hydrodynamical simulations of giant planets in low viscosity discs 
\citep{valborro07,lin10,lin11a,lin12,zhu13}. The fact that this is due to
the RWI has been explicitly verified through linear stability
analysis 
\citep{valborro07,lin10}. Here, we simulate gap-opening giant planets
in 3D discs where the kinematic viscosity varies with height above the
disc midplane. Our numerical setup is similar to
those that in \cite{pierens10}, but our interest is gap
stability in a layered disc. 
 
\subsection{Radially smooth viscosity profile for disc-planet
  interaction}\label{planet_visc_mode} 
\begin{figure}
  \centering
  \includegraphics[width=\linewidth]{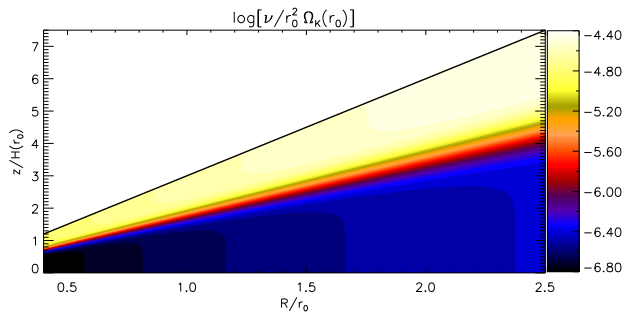}
  \caption{Example of the viscosity profile
    imposed in disc-planet simulations
    (Eq. \ref{planet_visc_profile}).  
    For a disc with constant aspect ratio, the viscous
    layer occupies a constant number of scale heights across the
    radial range. This specific plot corresponds to case P1, so the
    viscous layer (yellow-white colours) always occupies the uppermost
    $H$ at each cylindrical radius. 
    The solid line delineates the upper boundary of the computational
    domain. 
    \label{planet_visc2d}}
\end{figure}

Using the same notation as \S\ref{visc_model}, we impose a viscosity
profile $\hat{\nu}$ such that 
\begin{align}\label{planet_visc_profile}
  \hat{\nu}\Sigma_i(R)=
  \hat{\nu}_0\left[1+Q(\psi)\right]\Sigma_i(r_0)   
\end{align}
We have set the dimensionless argument in Eq. \ref{step} to
$\zeta=\psi$. Recall $\psi=\pi/2-\theta$ is the angular height away from the midplane. 
Viscosity increases from its floor value $\hat{\nu}_0$ by a factor
$A_\nu$ for $\psi > \zeta_\nu$. So the viscous layer is 
a wedge in the meridional plane, which conveniently fits into our
spherical grid.  
The angular thickness of the viscosity
transition is fixed to $\Delta\zeta_\nu =
0.2h$. Fig. \ref{planet_visc2d} gives an example of this  
viscosity profile. 

\subsection{Disc-planet simulations} 
We simulate locally isothermal discs with constant aspect-ratio
$h=0.05$ (by choosing $q=1$), vertical extent $n_h=3$ scale-heights 
and radial extent $[\rin,\rout]=[0.4,2.5]r_0$. Initially the surface density is smooth
($A=1$) with zero meridional velocity ($v_r=v_\theta=0$). 
The standard resolution is $(N_r, N_\theta,
N_\phi)=(256, 96, 768)$, corresponding to $6,\,32,\,6$ 
cells per $H$ along the $r,\theta,\phi$ directions at the reference
radius. We apply a damping rate $\hat{\gamma}=2$ with the reference
velocity field $\bm{v}_\mathrm{ref}=(0,0,v_\phi)$ in spherical
co-ordinates.   

\begin{figure}
  \centering
  \includegraphics[width=\linewidth]{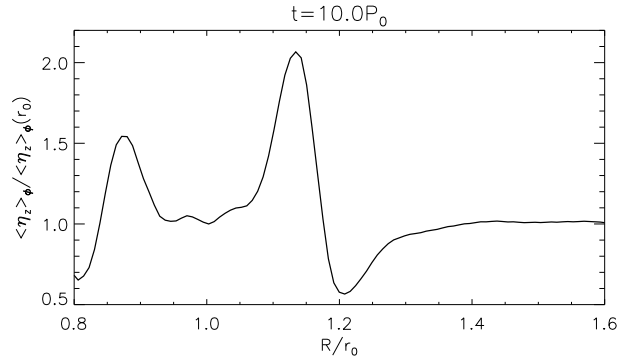}
  \caption{  Potential vorticity profile for a planet gap before
    it becomes unstable (case P0). The planet is located at $R=r_0$
    and the RWI first develops at the PV minima near the outer gap edge at
    $R\simeq1.2r_0$.
    \label{planet_PV}}
\end{figure}

We insert into the disc a planet of mass  
$M_p=10^{-3}M_*$, which corresponds to a Jupiter mass planet if
$M_*=M_{\sun}$. The softening length for the planet potential is
$\epsilon_p=0.5r_h$. The planet potential is switched on 
smoothly over $t\in[0,10]P_0$. We note that the disc can be considered
as two-dimensional for gap-opening giant planets, because the Hill
radius $r_h$ exceeds the local scale-height $H$ ($r_h/H\simeq1.4$
in our cases).   Fig. \ref{planet_PV} shows a typical PV profile
  associated with the gap induced by the planet. 

We remark that, apart from the viscosity prescription, the above
choice of physical and numerical parameter values are typical for
global disc-planet simulations \citep[e.g.][]{valborro06,mignone12}.    

\subsection{Results}
Table \ref{planet_sims} summarizes the disc-planet simulations. 
The main simulations to be discussed are cases P0 --- P1, with a floor
viscosity of $\hat{\nu}_0=2.5\times10^{-7}$. 
The fiducial run P0 has $A_\nu=1$, i.e. no viscous layer, so that $\alpha\sim 10^{-4}$ everywhere. 
The more typical viscosity value adopted for disc-planet simulations,
$\hat{\nu}\sim 10^{-5}$ or $\alpha\sim 10^{-3}$, is known to suppress vortex formation
\citep{valborro07, mudryk09}. Thus vortex formation is expected in 
case P0. For case P0.5 and P1 we set $A_\nu=100$ with transition angle $\zeta_\nu=2.5h$ and
$2h$, respectively, so the viscous layer with $\alpha\sim10^{-2}$ occupies the uppermost $0.5H$ and $H$ of the vertical
domain. Case P0R is case P0 restarted from 
$t=100P_0$ with the layered viscosity profile of case P1.

\begin{table*}
  \centering
  \caption{Summary of disc-planet simulations. These runs employ the
    `wedge' viscosity model described by
    Eq. \ref{planet_visc_profile}. The thickness of the viscous layer
    is measured from the upper disc boundary. The $m=1$ mode amplitude was 
    averaged over the shell $r\in[1.2,1.6]r_0$; and the overbar
    denotes a further time average over    
    $t\in[t_\mathrm{max},200]P_0$, where $t_\mathrm{max}$ is when
    $\mathrm{max}(a_1)$ is attained. Case P0R employs the
    viscosity profile of P0 for $t\leq100P_0$, and that of P1 for 
    $t>100P_0$.\label{planet_sims}}
    \begin{tabular}{llllrrl}
      \hline\hline
      Case & $10^6\hat{\nu}_0$ & $A_\nu$ & visc. layer&
      $10^2\overline{a}_1$&$10^2a_1(200P_0)$ & comment \\ 
      \hline
      P0     & 0.25  & 1            & 0     & 18.3  & 14.5 & single vortex
       by $t=130P_0$ and persists until end of sim.  \\ 
      P0.5   & 0.25  & 100          & $0.5H$ &  17.4  & 8.6& single vortex
      by $t=90P_0$ and persists until end of sim.\\ 
      P0R    & 0.25  & 1$\to$100    & $0\to H$& 10.1 & 2.5& single vortex
      by $t=130P_0$, disappears after $t=180P_0$ \\
      P1     & 0.25  & 100          & $H$    & 12.5  & 2.1 &single vortex
      by $t=80P_0$, disappears after $t=170P_0$ \\   

      Pb0     & 1.0  & 1          & 0      & 9.2  & 5.4 &single vortex by $t=130P_0$, disappears
       after $t=160P_0$   \\ 
      Pb0.5   & 1.0  & 10         & $0.5H$ & 7.3  & 3.0 &similar to Pb0     \\ 
      Pb1     & 1.0  & 10         & $H$    & 5.7  & 2.2   &single vortex
      by $t=120P_0$, disappears after $t=140P_0$   \\ 

      Pc0.5   & 1.0  & 100          & $0.5H$ & 4.2  &  2.3 &two vortices
      at $t\sim100P_0$, no vortices after $t\sim120P_0$   \\ 
      Pc1     & 1.0  & 100          & $H$    & 2.2 &  1.6 &two weak vortices at
      $t\sim60P_0$, no vortices after $t\sim 80P_0$\\ 
      \hline
  \end{tabular}
\end{table*}


\subsubsection{Density evolution}

\begin{figure*}
  \centering
  \includegraphics[scale=.43,clip=true,trim=0cm 1.84cm 0cm
    0cm]{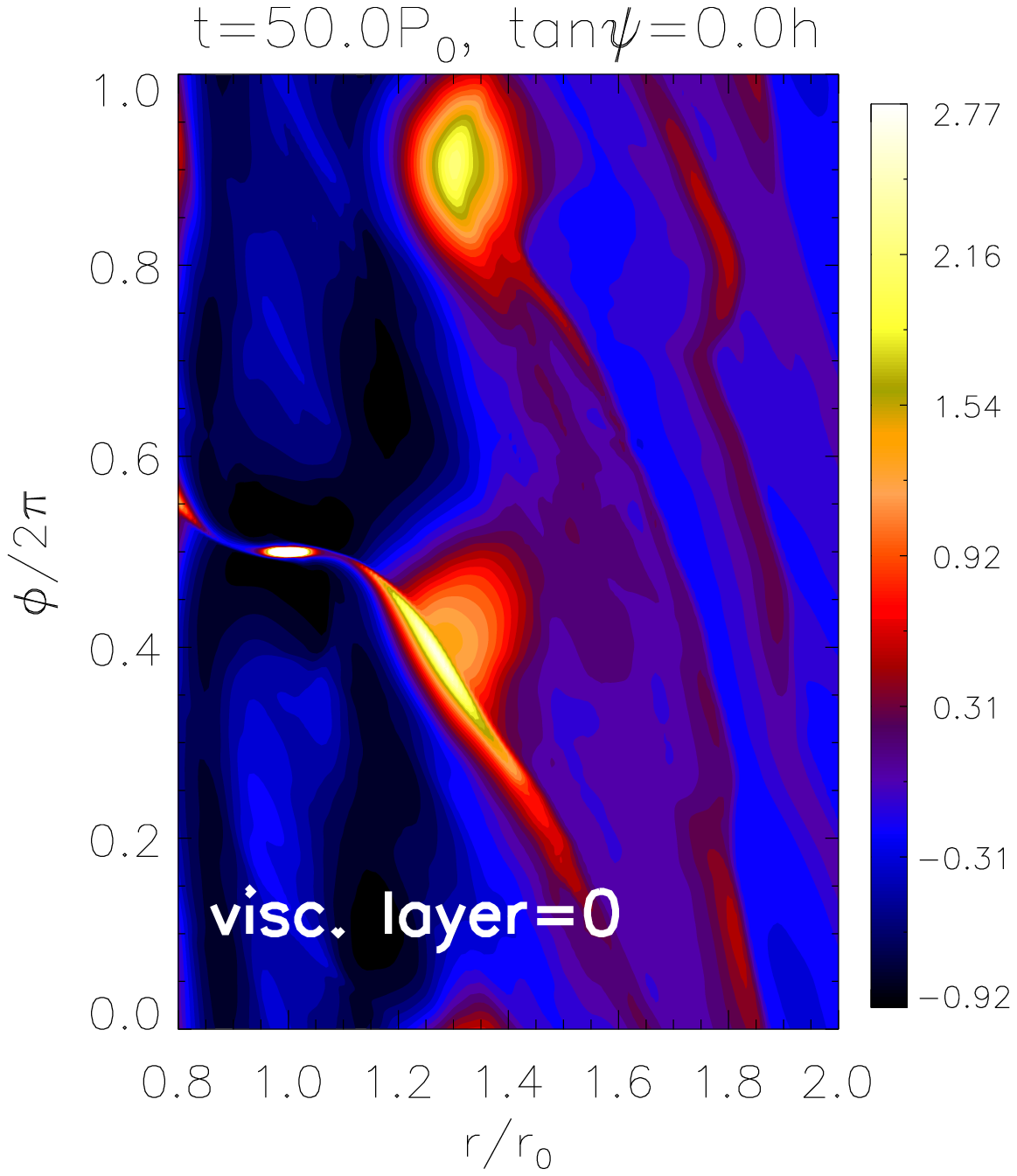}\includegraphics[scale=.43,clip=true,trim=2.3cm
    1.84cm 0cm 0cm]{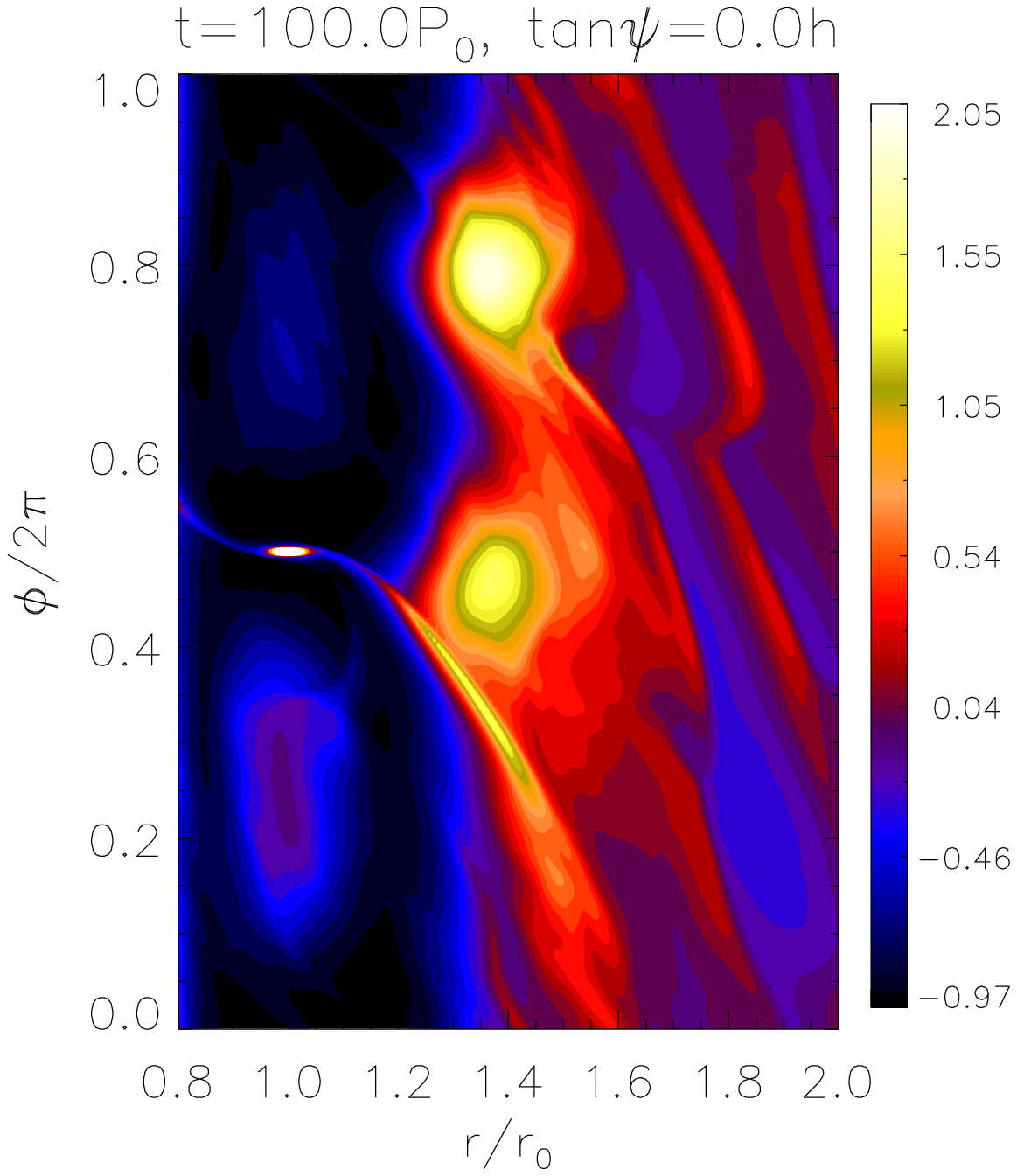}\includegraphics[scale=.43,clip=true,trim=2.3cm
    1.84cm 0cm 0cm]{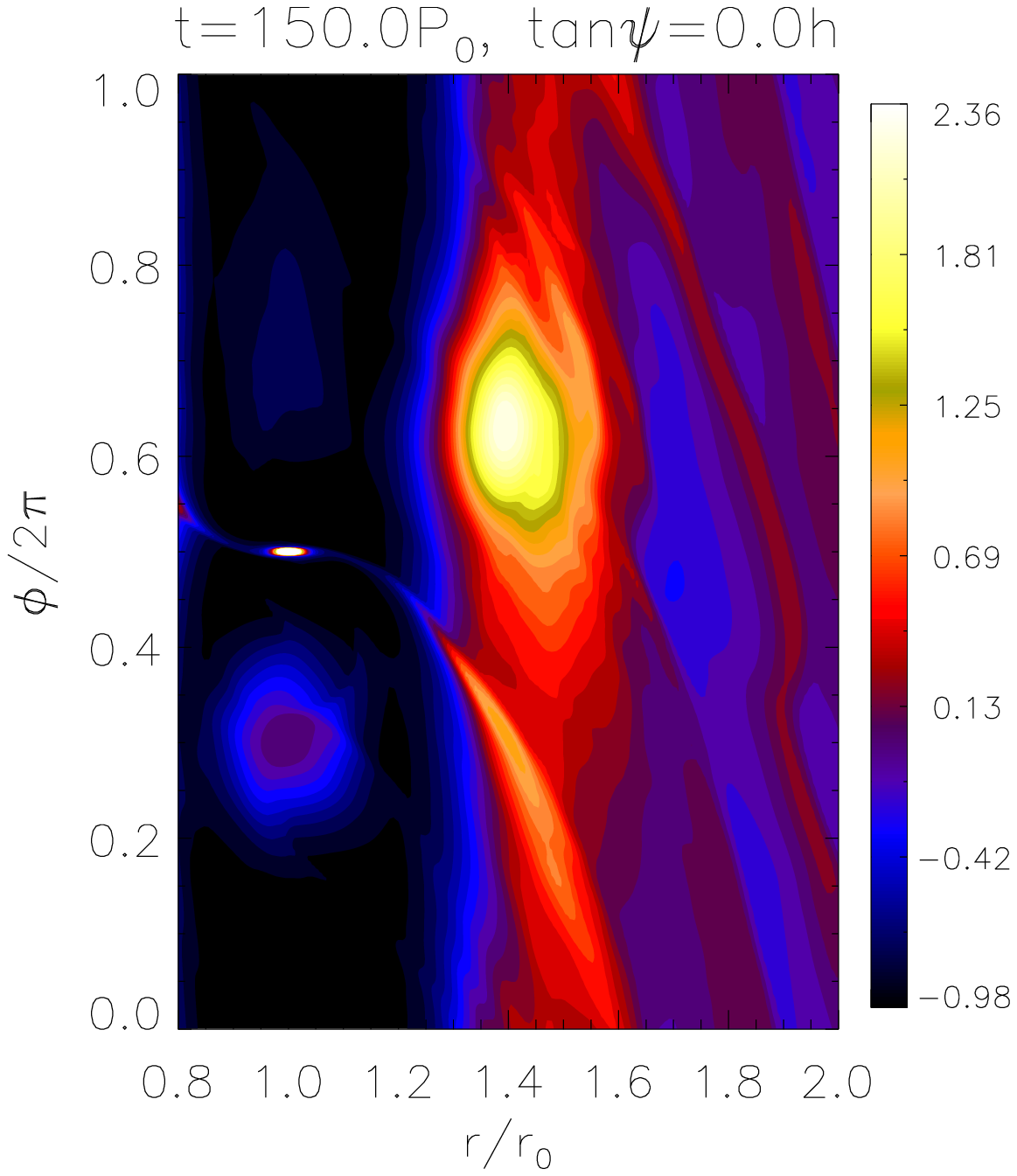}\includegraphics[scale=.43,clip=true,trim=2.3cm
    1.84cm 0cm 0cm]{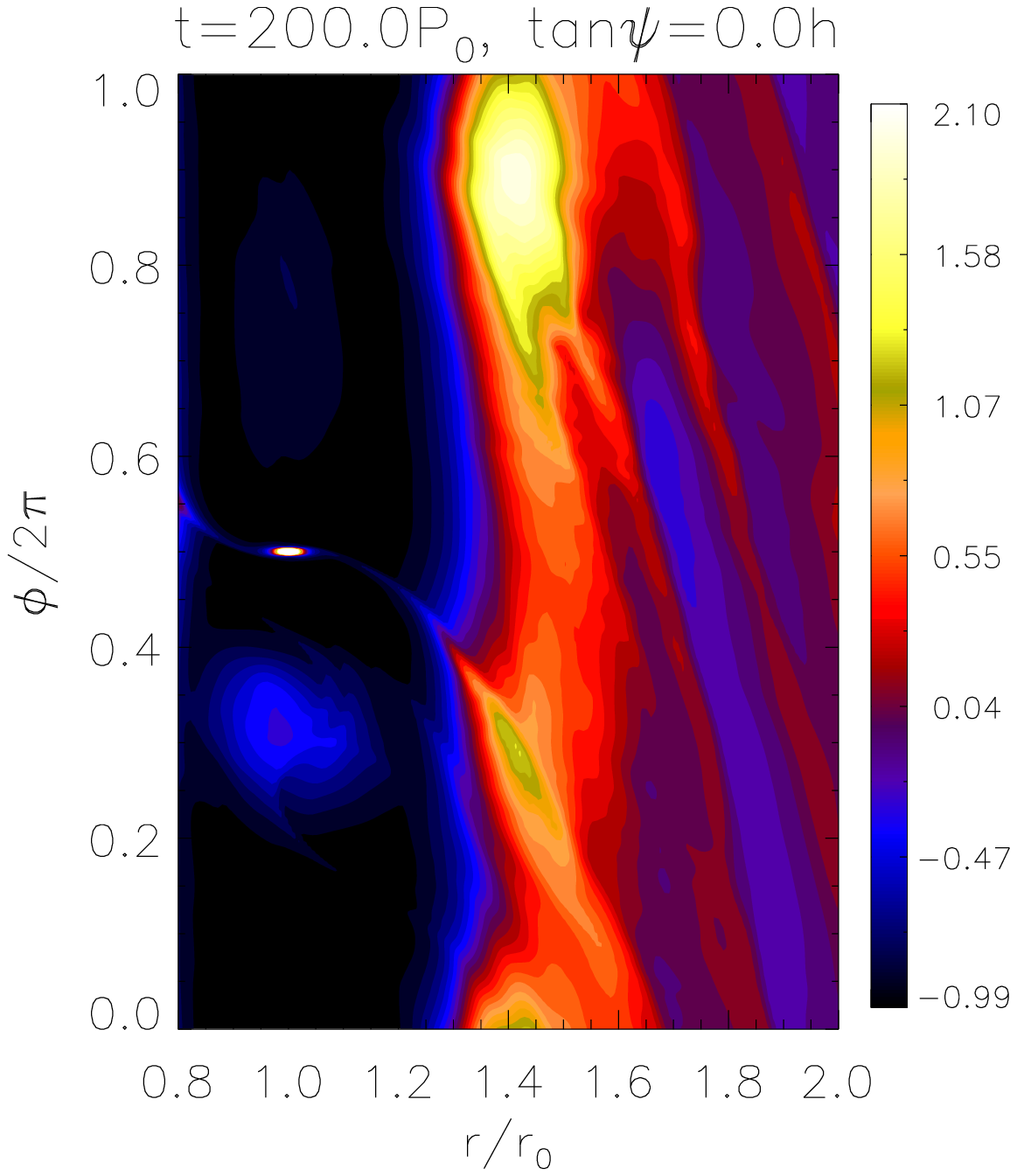}\\
  \includegraphics[scale=.43,clip=true,trim=0cm 1.84cm 0cm
    0.9cm]{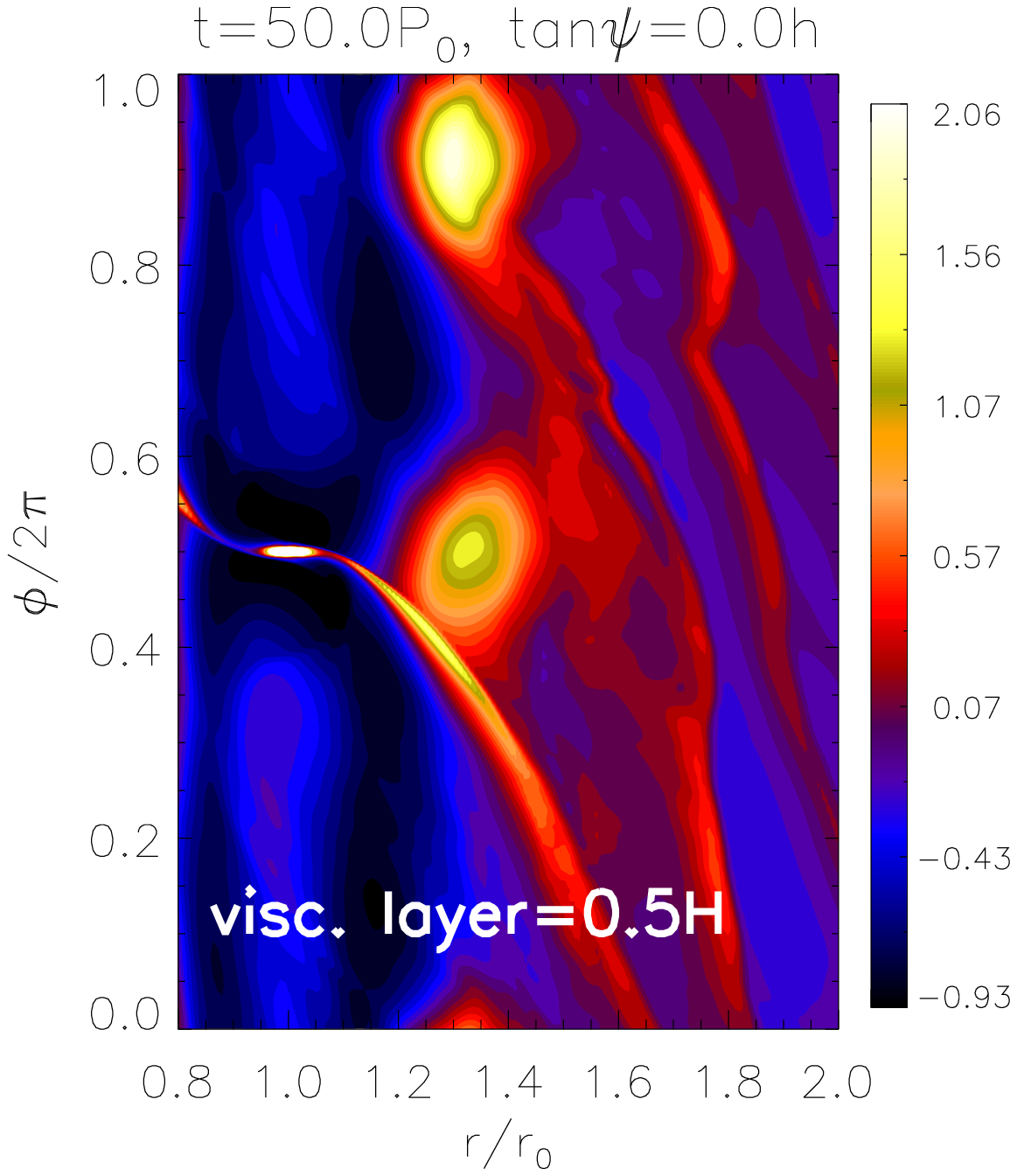}\includegraphics[scale=.43,clip=true,trim=2.3cm
    1.84cm 0cm 0.9cm]{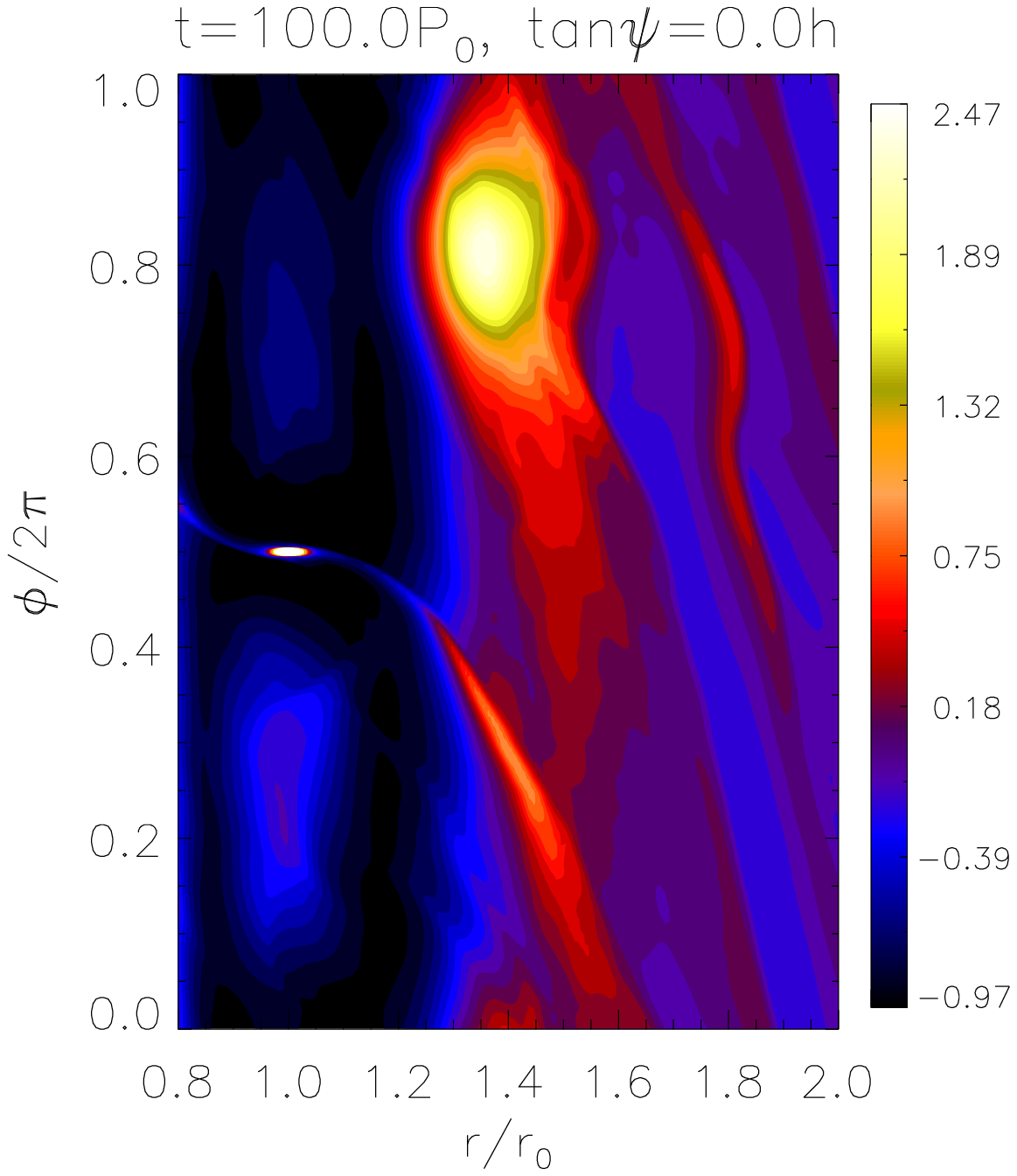}\includegraphics[scale=.43,clip=true,trim=2.3cm
    1.84cm 0cm 0.9cm]{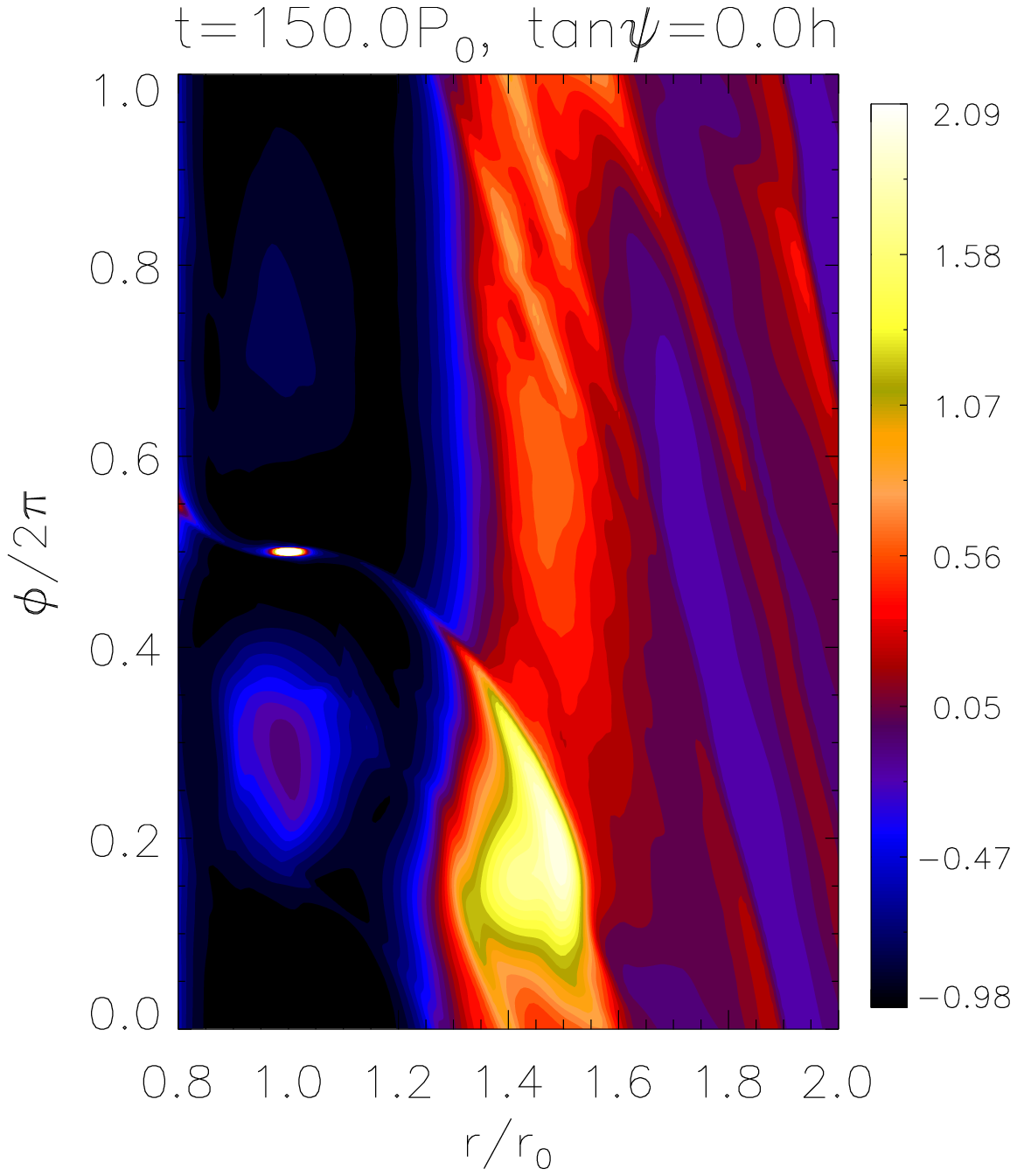}\includegraphics[scale=.43,clip=true,trim=2.3cm
    1.84cm 0cm 0.9cm]{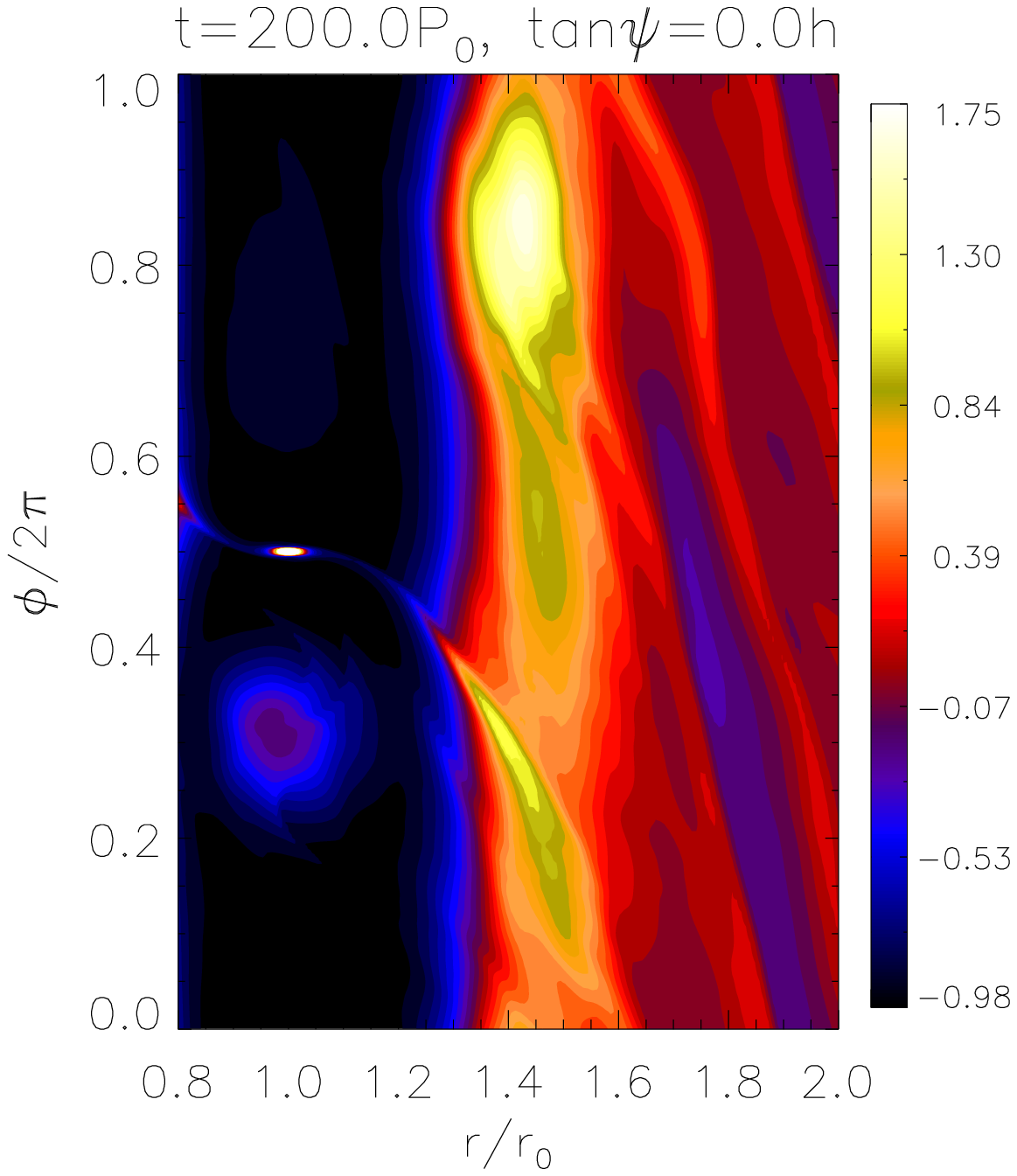}\\
  \includegraphics[scale=.43,clip=true,trim=0cm 0.cm 0.cm
    0.9cm]{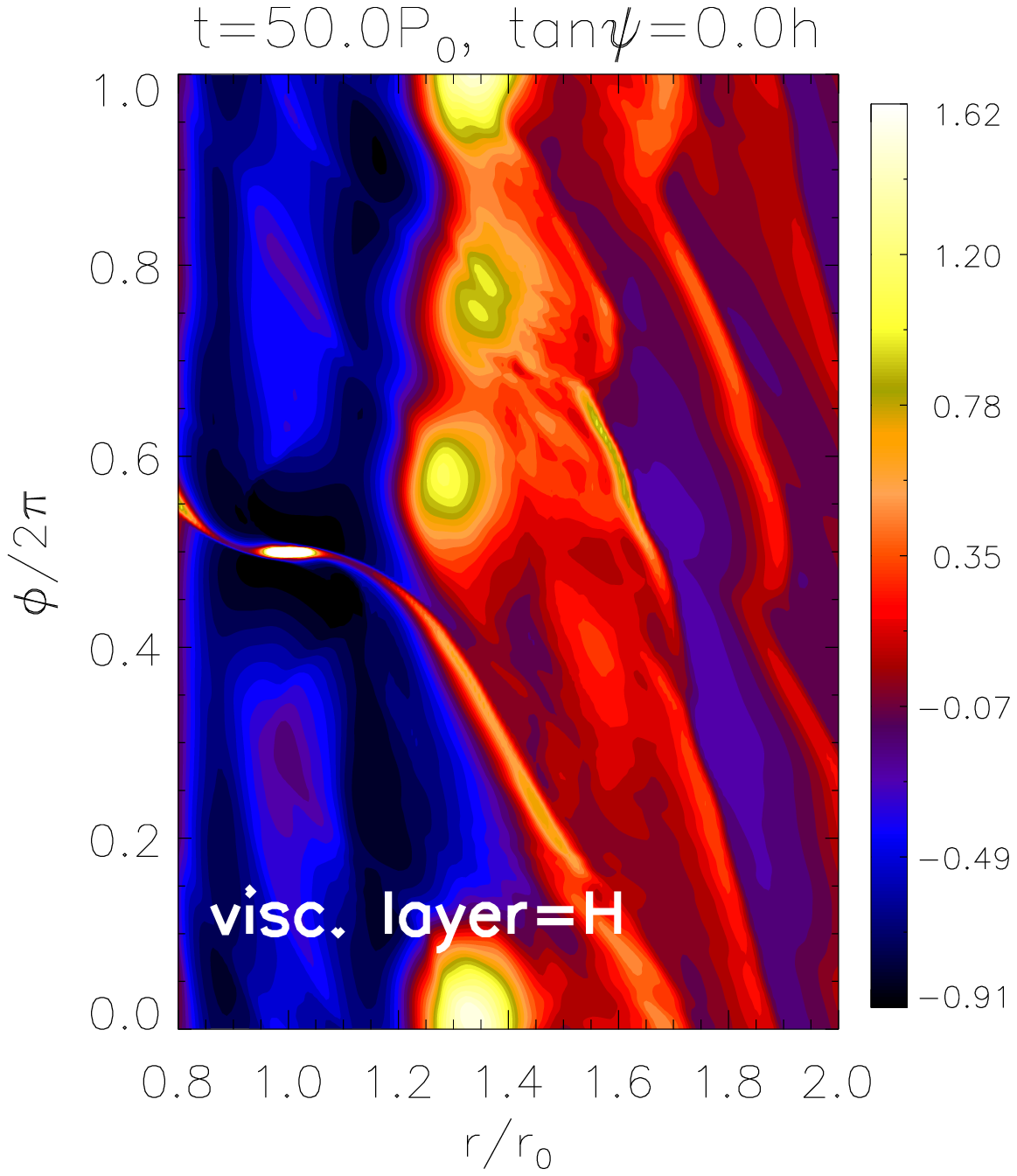}\includegraphics[scale=.43,clip=true,trim=2.3cm
    0.cm 0cm 0.9cm]{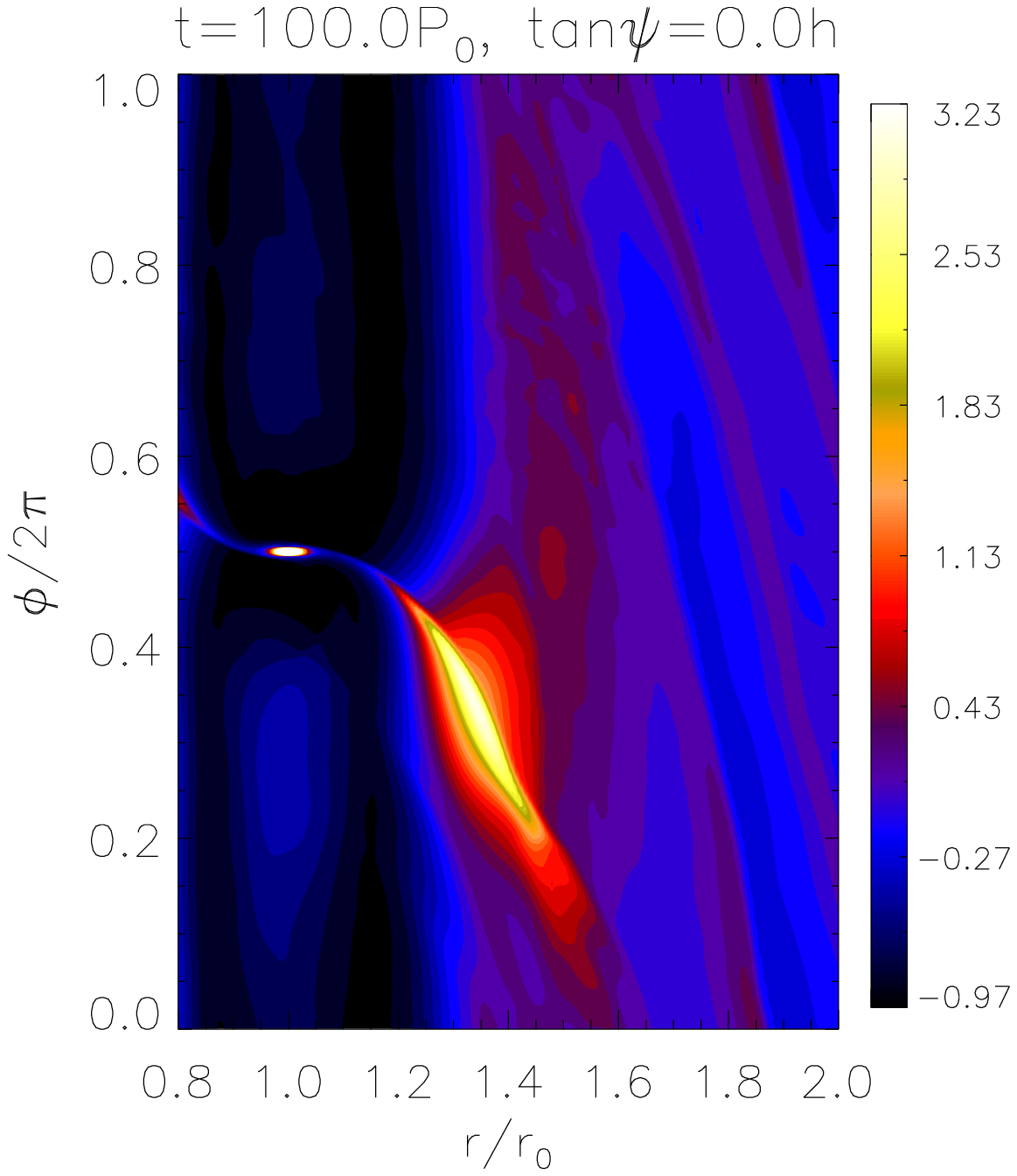}\includegraphics[scale=.43,clip=true,trim=2.3cm
    0.cm 0cm 0.9cm]{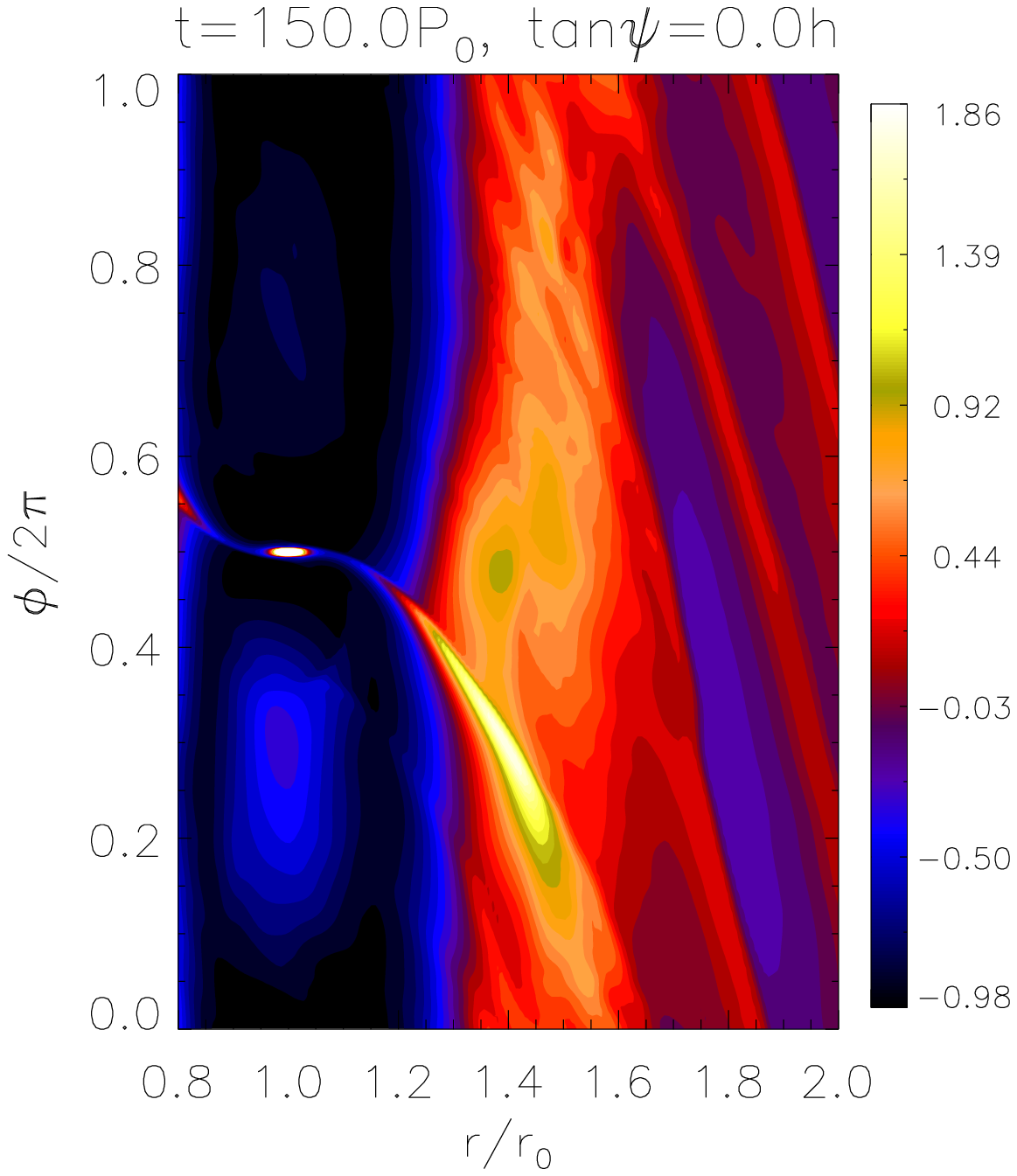}\includegraphics[scale=.43,clip=true,trim=2.3cm
    0.cm 0cm 0.9cm]{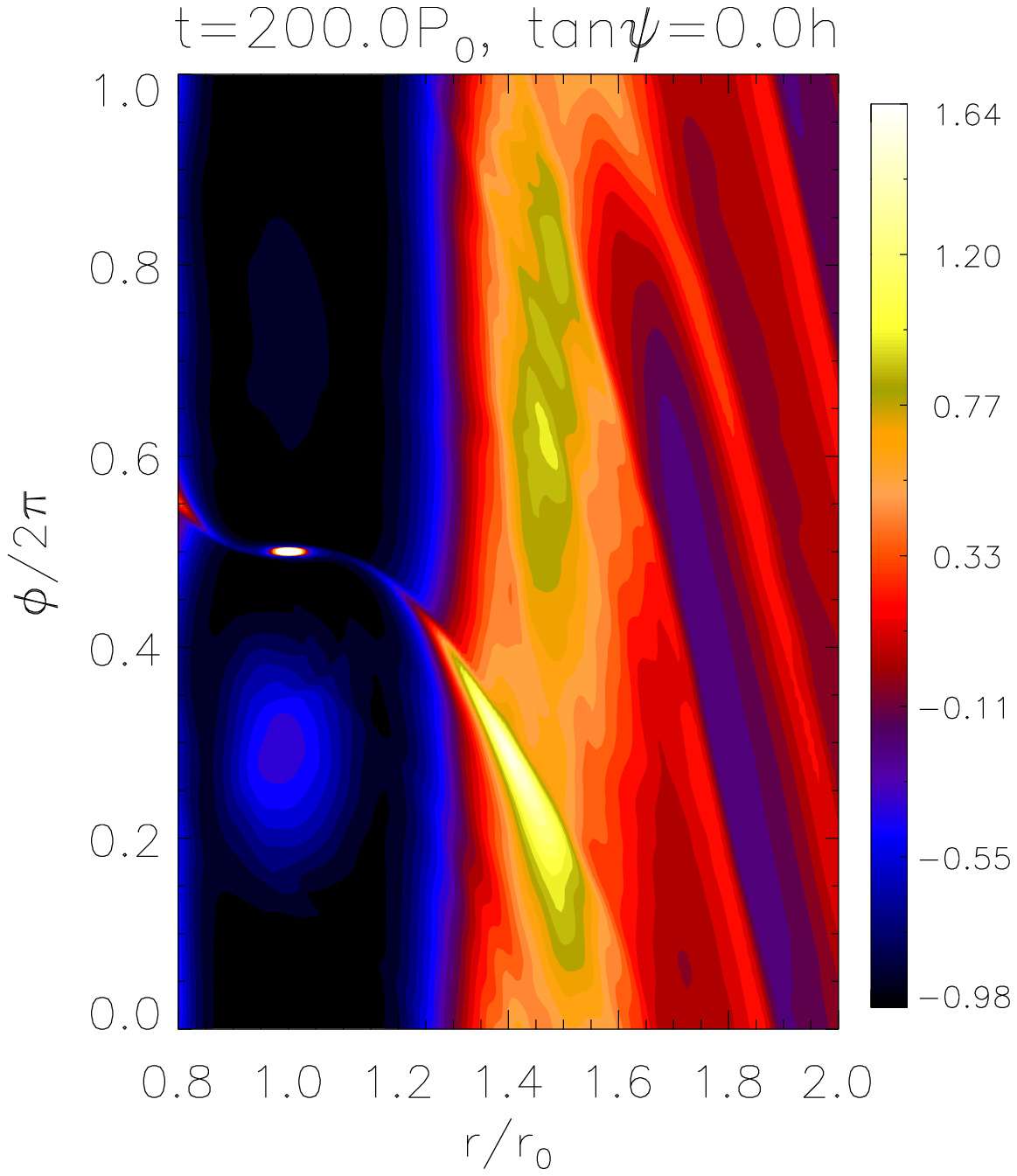}\\
  \caption{Relative density perturbation $\delta\rho$ for disc-planet
    simulations. Top: case P0 (no viscous layer), middle: case P0.5
    (viscous layer of $0.5H$), bottom: case P1 (viscous layer of $H$). The
    vertical extent of the computational domain is $3H$ and the
    viscous layer is measured from the upper disc boundary.  
    \label{jup0_3h}}
\end{figure*}

Fig. \ref{jup0_3h} compares the time evolution of the midplane density
perturbation $\delta\rho(z=0)$ for cases P0, P0.5 and P1. In 
all cases we observed the RWI with $m=4$ develops early on ($t\simeq20P_0$),
consistent with the limited effect of viscosity on the linear
instability, as found above. 
The no-layer case P0 and layered case P0.5 (viscous layer of $0.5H$)
behave similarly, showing that a thin viscous layer has little effect
on the evolution of the unstable gap edge, at least over the
simulation time-scale of $200P_0$.  


Case P1 evolves quite
differently from case P0. While a single vortex does form at
$t\sim100P_0$, it is \emph{transient}, having disappeared at the end of
the simulation for P1. The final $m=1$ amplitude is about 3 times smaller
than that in case P0 (Table \ref{planet_sims}). This result is
significant because the upper viscous layer in case P1, of 
thickness $H$, only occupies $\sim4\%$ of the total column density, 
but the vortex is still destroyed. This suggests that vortex
survival at planetary gap edges requires low effective viscosity
throughout the vertical fluid column. 

\subsubsection{Kinetic energy density}
\begin{figure}
  \centering
  \includegraphics[width=\linewidth]{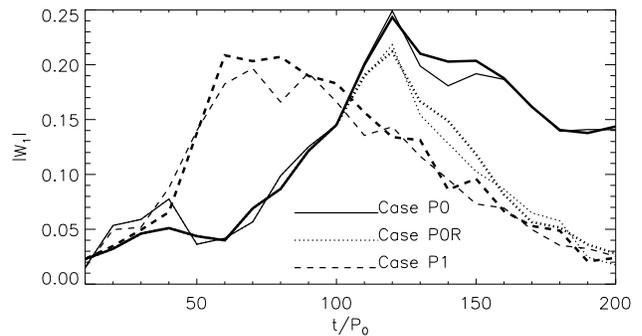}
  \caption{Evolution of the $m=1$ component of kinetic energy density,
    averaged over $r\in[1.2,1.6]r_0$. This average
    is split into that taken over disc bulk ($z\in[0,2]H$, thick
    lines) and the disc atmosphere ($z\in[2,3]H$, thin lines). Case P0
    has no viscous layer (solid) and case P1 has a viscous layer in
    $z\in[2,3]H$ (dashed). Case P0R (dotted) is identical to P0 up to $t=100P_0$,
    but was simulated for $t>100P_0$ with a viscous layer of thickness
    $H$.  
\label{pdisk_kerz_cases_planet}}
\end{figure}

Here, we compare the $m=1$ component of the kinetic energy density
($W_1$)  between the no-layer case P0, layered
case P1 and case P0R which is P0 resumed from $t=100P_0$ with a
viscous layer. Fig. \ref{pdisk_kerz_cases_planet} shows 
$W_1(t)$ averaged over the outer gap edge. For each case we
average $W_1$ over the disc bulk and the atmosphere, and plot them
separately in the figure. 

The $m=1$ component does not emerge from the linear instability, but is a
result of non-linear vortex merging. 
Fig. \ref{pdisk_kerz_cases_planet} shows that merging is accelerated
by a viscous layer: the single vortex appears at $t\sim70P_0$ for case
P1 but only forms at $t\sim120P_0$ for case P0. Also note for all
cases, $W_1$ in the disc bulk (thick lines) is similar to that in the
disc atmosphere (thin lines), implying the $m=1$ disturbance
(i.e. the vortex) evolves two-dimensionally. We checked that this is
consistent with the Froude number $Fr\equiv|Ro|H/z < 1 $ away from the
midplane \citep{barranco05,oishi09}. 


Case P0R shows that introducing a viscous layer eventually destroys
the vortex. The local viscous timescale is $t_\nu\equiv
H^2/\nu\gtrsim 16P_0$, so on short timescales after introducing the 
viscous layer ($t=100P_0$), vortex-merging proceeds in case P0R
similarly to case P0 ($t\in[100,110]P_0$). However, $W_1$ decays for
$t>110P_0$ and evolves towards that of case P1. We expect viscosity
to damp the $m=1$ disturbance in the disc atmosphere between
$t\in[110,200]P_0$ because this corresponds to $\sim 6 t_\nu$, but 
the disturbance in the disc bulk is also damped out: the evolution
remains two-dimensional.    

We emphasize the kinetic energy is dominated by horizontal
motions, with $\mathrm{max}(|v_z|/|\bm{v}|) < 0.03$ at the outer
gap edge ($r\in[1.2,1.6]r_0$). Vertical motions are well sub-sonic. When averaged
over $z\in[0,2]H$ and $z\in[2,3]H$, the vertical Mach number
$M_z\equiv|v_z|/c_s \simeq 0.05,\,0.08$ (P0), $M_z\simeq 0.04,\,0.06$ (P0R) and
$M_z \simeq 0.05,0.06$ (P1), respectively. 


\subsubsection{Potential vorticity}
\begin{figure*}
  \centering
  \includegraphics[scale=.43,clip=true,trim=0cm .0cm 0cm
    0cm]{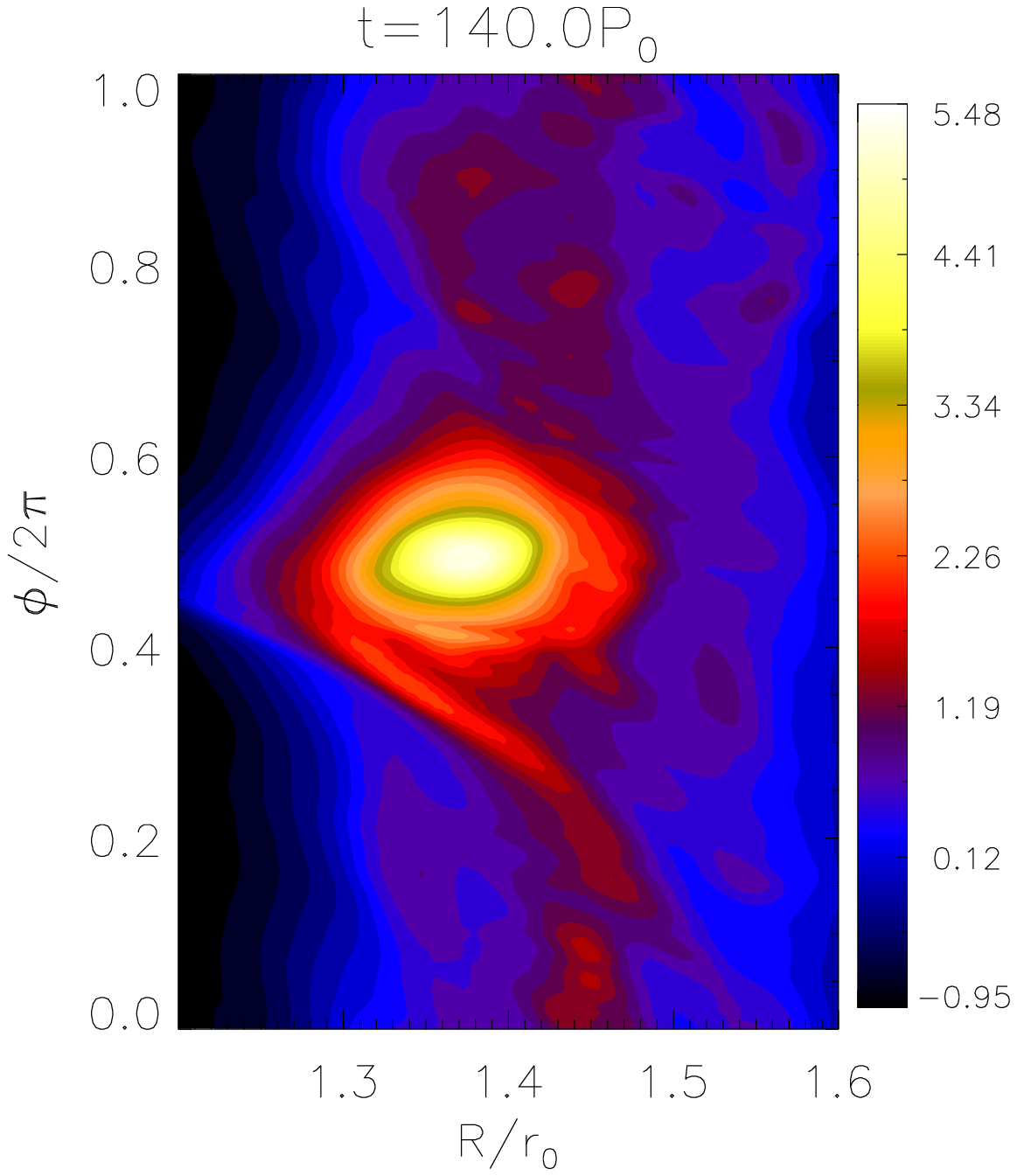}\includegraphics[scale=.43,clip=true,trim=2.3cm
    .0cm 0cm 0cm]{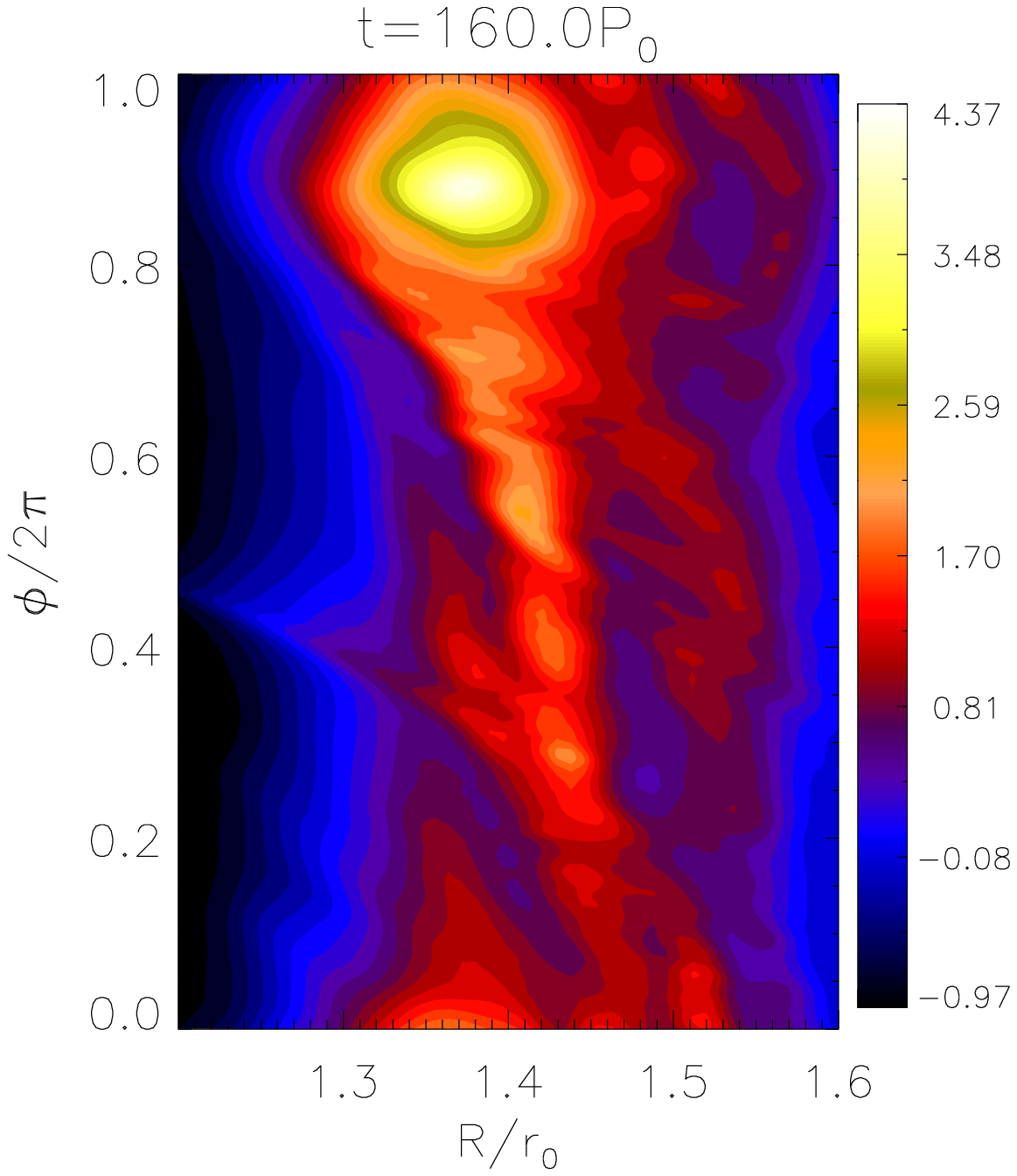}\includegraphics[scale=.43,clip=true,trim=2.3cm
    .0cm 0cm 0cm]{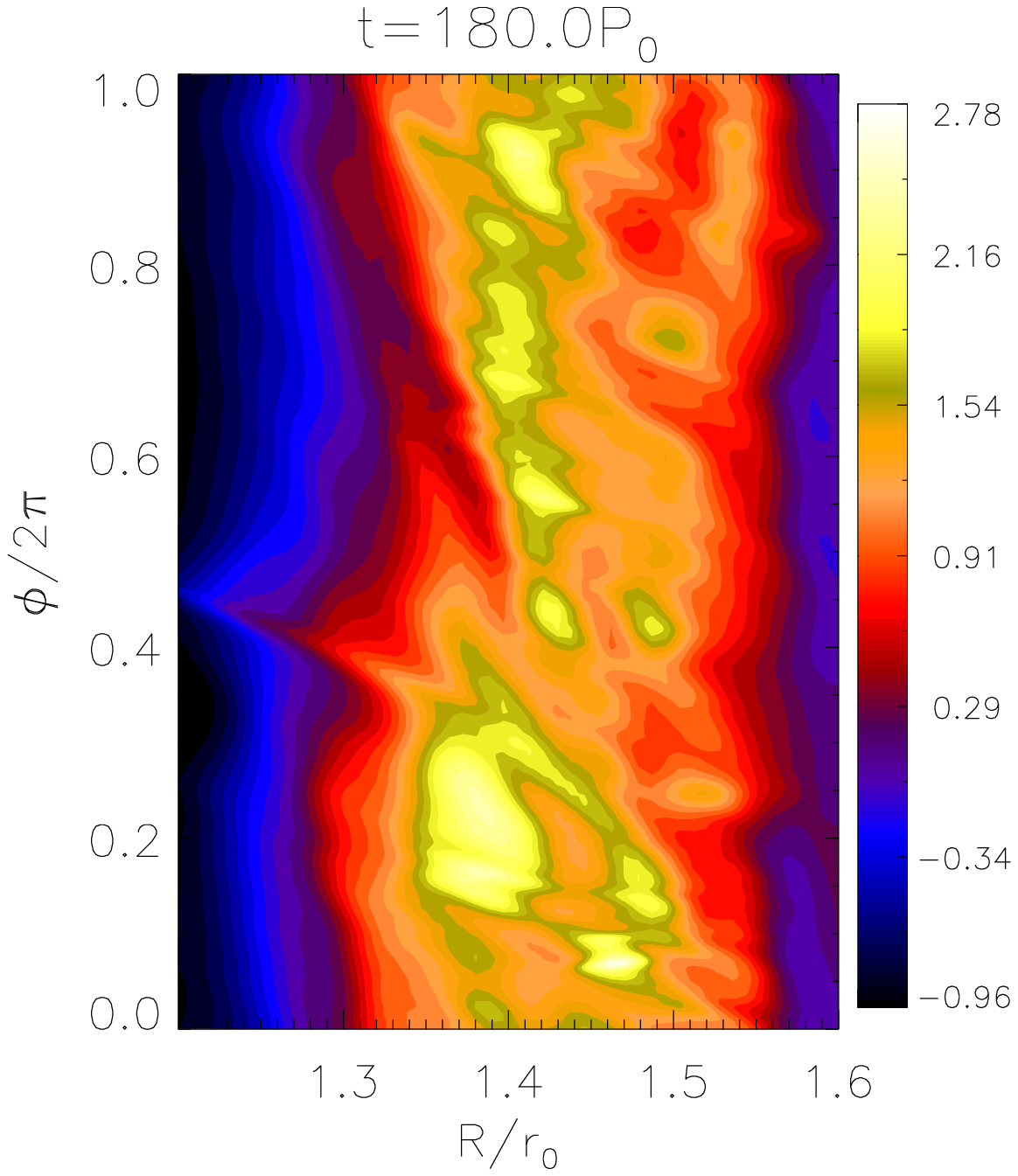}\includegraphics[scale=.43,clip=true,trim=2.3cm
    .0cm 0cm 0cm]{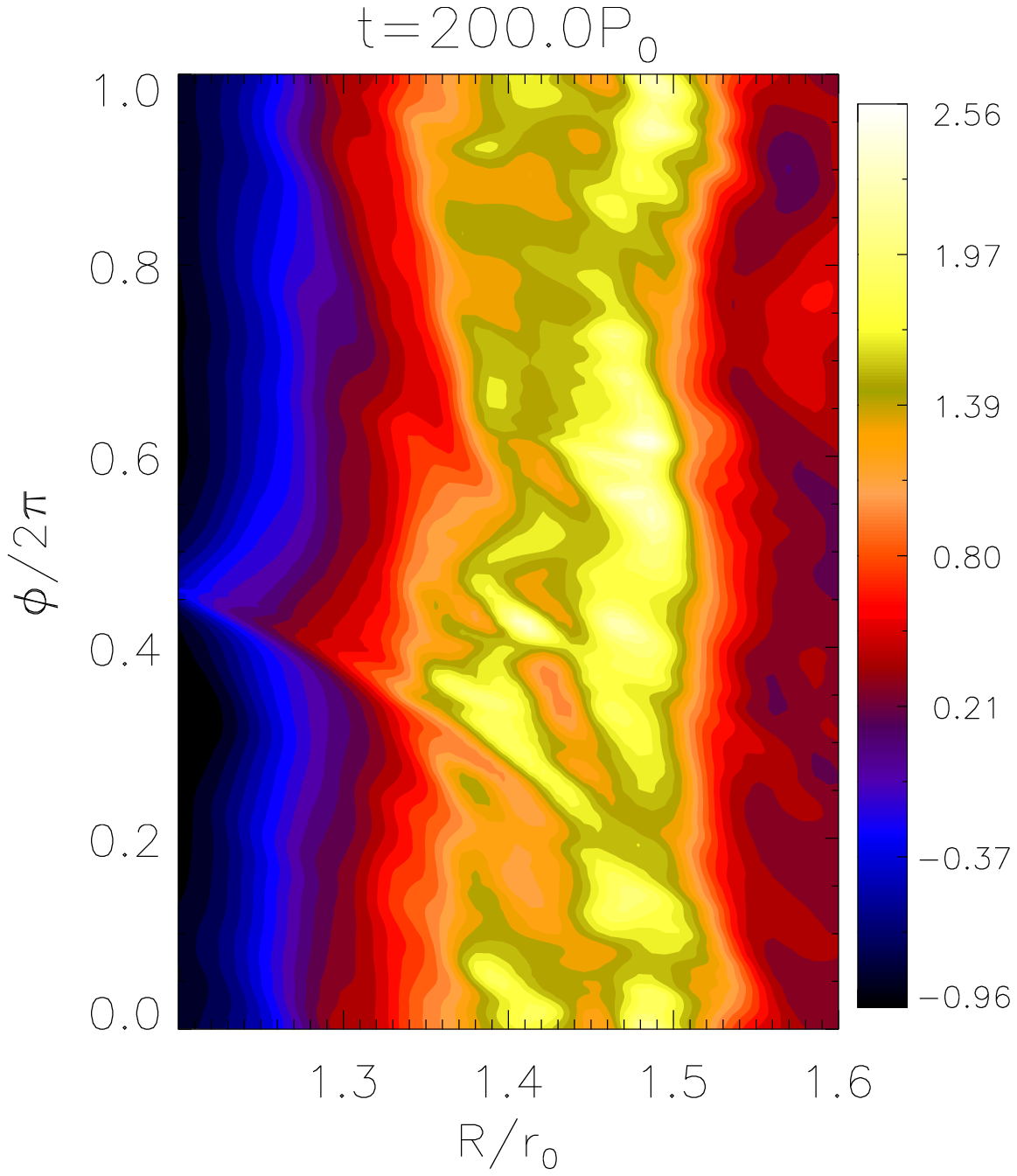}
  \caption{Inverse PV perturbation, $\eta_z(t=0)/\eta_z - 1$, for case P0R, which was resumed
    from the no-layer case P0 from $t=100P_0$ with the introduction of
    a viscous layer of $H$. 
    \label{jup0_3h_visc_restart_vorten}}
\end{figure*}

We examine the PV evolution for case P0R in
Fig. \ref{jup0_3h_visc_restart_vorten}.  
To highlight the vortices, which are positive (negative)
density (vertical vorticity) perturbations, we show the inverse PV
perturbation, $\delta\eta_z^{-1}\equiv\eta_z(t=0)/\eta_z - 1$. As
noted above, a single vortex still forms  
despite introducing a viscous layer at $t=100P_0$. However, it
decays rapidly compared to case P0. 
The region with $\delta\eta_z^{-1}>0$ (i.e. the vortex) elongates and 
shifts outward from $R\simeq 1.38r_0$ at $t=140P_0$ to $R\simeq1.5r_0$
at $t=200P_0$, by which the vortex has disappeared. (A similar
evolution was observed for case P1.) The vortex is stretched
azimuthally much more than radially. This is not surprising since the
imposed viscosity profile is axisymmetric. The important point is
that viscosity is only large near the disc surface, but still has a
significant effect on the vortex. 


\subsubsection{Resolution check}
We repeated simulations P0 and P1 with resolution
$(N_r,N_\theta,N_\phi)=(512,96,1536)$, corresponding
to $12$ and $32$ cells per scale-height in $(r,\phi)$ and $\theta$,
respectively. We denote these runs as P0HR and P1HR below. 

We observe similar evolution in P0HR and P1HR as their standard
resolution versions. However, due to lower numerical diffusion, we
find stronger vortices in P0HR. Although the vortex in P1HR persisted
longer than the standard resolution run, it was still subject to rapid
decay in comparison with P0HR. At $t=200P_0$ we find 
the $m=1$ amplitude to be $a_1=0.29$ and $a_1=0.10$ at the outer gap
edge, respectively for P0HR and P1HR; a similar contrast as that
between P0 and P1. A weak over-density was still observed in
P1HR at $t=200P_0$, but it further decays to $a_1=0.06$ at $t=230P_0$
and there is no vortex. By contrast, P0HR was simulated to $t=250P_0$
and the vortex survived with little decay ($a_1=0.25$).   
  
\begin{figure}
  \centering
  \includegraphics[scale=.375,clip=true,trim=0cm .0cm 0cm
    0cm]{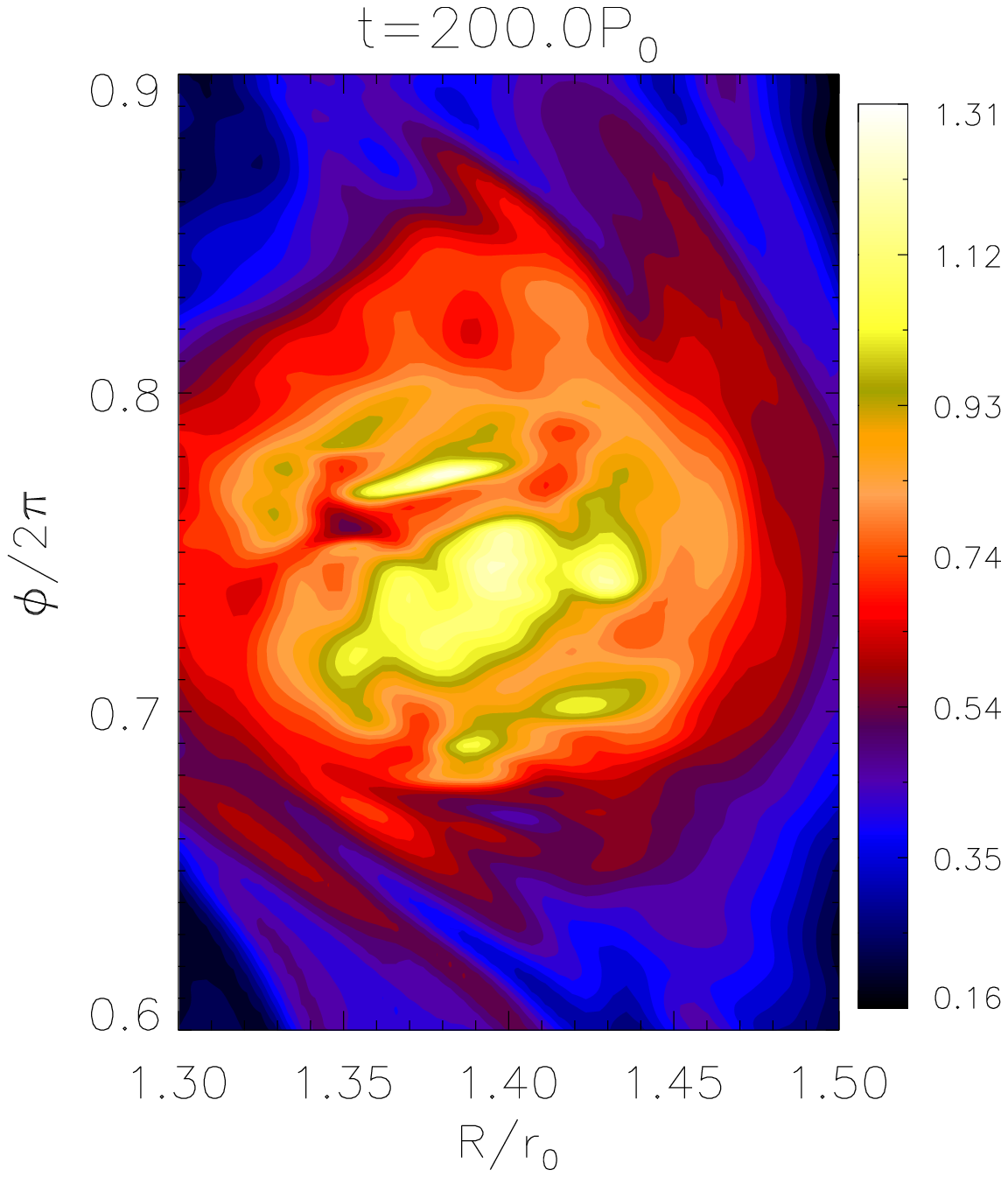}\includegraphics[scale=.375,clip=true,trim=1.7cm
    .0cm 0cm 0cm]{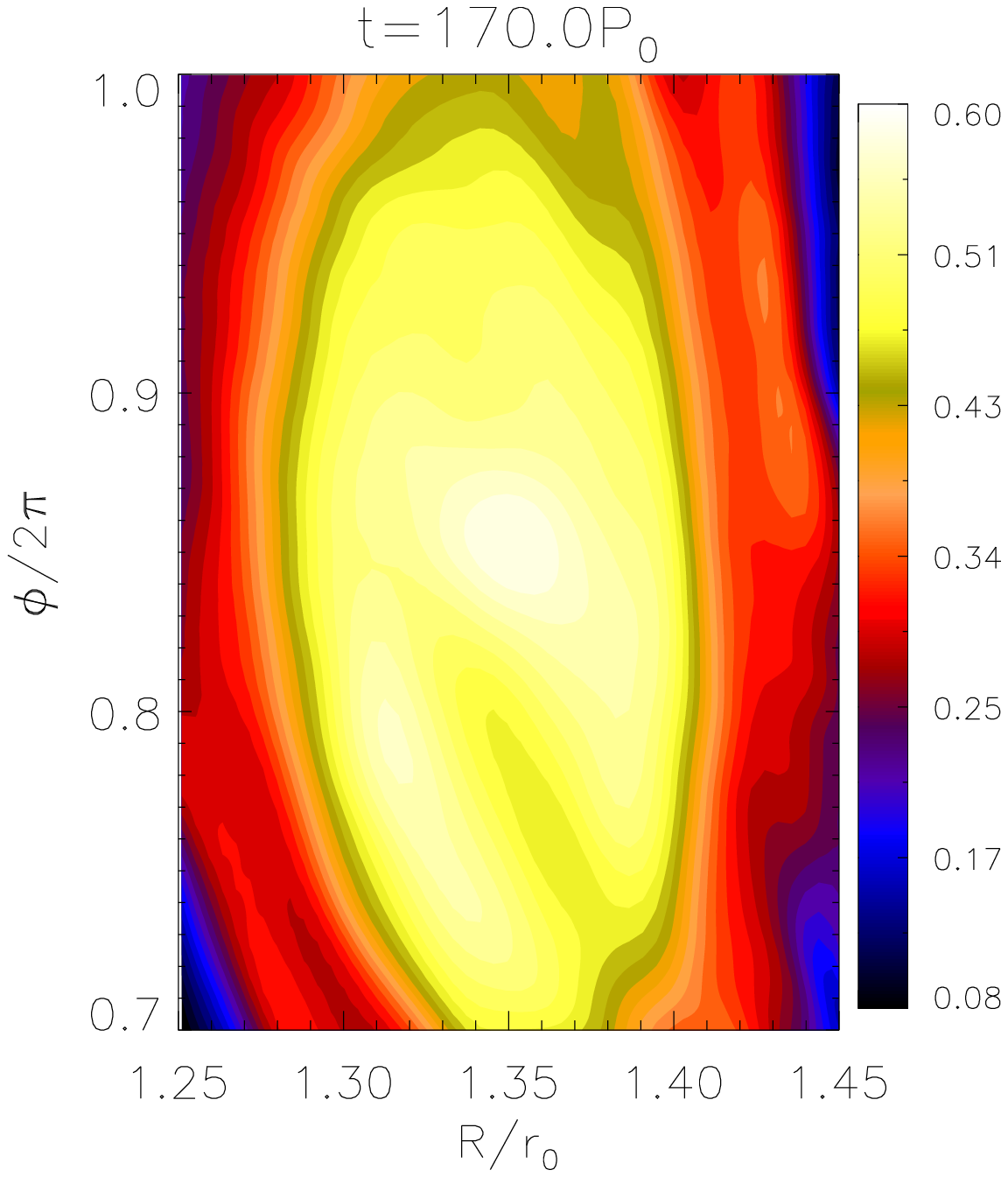}
  \caption{Logarithmic inverse PV perturbation associated with a
    vortex, $\log{\left[\eta_z(t=0)/\eta_z\right]}$, for high resolution cases
    P0HR (left, no viscous layer) and P1HR (right, with a viscous
    layer). These cases are the same as P0 and P1, but with double the
    $(r,\phi)$ resolution. The vortex in P1HR (right) eventually disappears
     after $t \sim 200P_0$. 
    \label{HR_sims}}
\end{figure}

Interestingly, we observe small-scale ($\lesssim H$) disturbances inside the
vortex in P0HR. This is shown in the left panel of  
Fig. \ref{HR_sims} in terms of the (inverse) PV perturbation. We
checked the density field remains smooth, so this small-scale structure is
due to vorticity variations. This is unlikely the elliptic instability
\citep{lesur09}, though, because the numerical resolution is still
insufficient for studying such instabilities; especially since the
vortex is  very elongated with large aspect ratio
$\sim 10$  \cite[so the elliptic instability is weak,][]{lesur09}.  
Despite the disturbances, the vortex over-density in P0HR remains coherent until the
end of the simulation, possibly because the planet maintains the 
condition for RWI. On the other hand, the vortex in the layered-case
P1HR does not develop small-scale structure (Fig. \ref{HR_sims}, right
panel), yet it is destroyed by the end of the simulation.  

\subsection{Additional simulations}
Locally isothermal, low viscosity discs are vulnerable to the
so-called `vertical shear 
instability' because $\partial_z\Omega_i\neq 0$ \citep{nelson12}. 
\citeauthor{nelson12} employed a radial  resolution $\gtrsim 60$ cells
per $H$ to resolve this instability because it involves small radial
wavelengths ($\ll H$). Our numerical resolution is unlikely to capture
this instability. Nevertheless, we have performed additional
simulations designed to eliminate the vertical shear instability.   

\subsubsection{Larger floor viscosity}
We performed several simulations with $\hat{\nu}_0=10^{-6}$. A viscosity of
$\hat{\nu}\sim 10^{-6}$ is expected to damp the vertical shear 
instability \citep{nelson12}, while still permitting the gap-edge
RWI. Table \ref{planet_sims} summarizes
these cases with $A_\nu=10$ (`Pb' runs) and $A_\nu=100$ (`Pc' runs). 

\begin{figure}
  \centering
  \includegraphics[width=\linewidth]{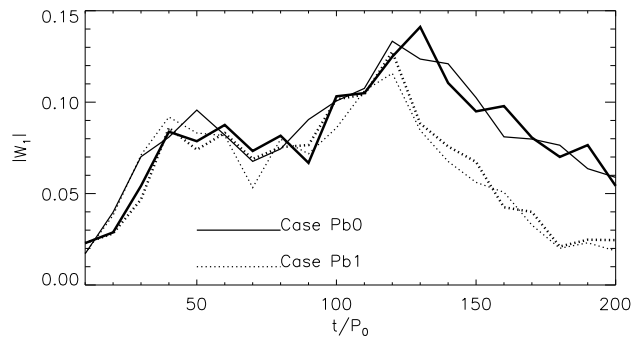}
  \caption{Same as Fig. \ref{pdisk_kerz_cases_planet} but with
    floor viscosity $\hat{\nu}_0=10^{-6}$: cases Pb0 (solid, no viscous layer) and Pb1
    (dotted, viscous layer of $H$). The thick (thin) lines indicate
    $W_1$ averaged over $z\in[0,2]H$
    ($z\in[2,3]H$). 
    \label{pdisk_kerz_cases_planet_hivisc}}
\end{figure}

In these simulations we find vortices eventually decay, even in the
no-layer case Pb0. For $A_\nu=10$, the layered cases Pb0.5 and Pb1 evolve
similarly  to Pb0: three vortices formed by $t\sim30P_0$, merging into two
vortices by $t\sim40P_0$, then finally into a single vortex by
$t\sim130P_0$, which subsequently decays. However, the final vortex
decays faster in the presence of a viscous layer. This is shown in
Fig. \ref{pdisk_kerz_cases_planet_hivisc}, which compares the $m=1$
kinetic energy density for case Pb0 and Pb1. The evolution only begins
to differ after the single-vortex has formed. 

For $A_\nu=100$ (cases Pc0.5 and Pc1), we find the $m=2$ amplitude
dominated over $m=1$, so a single-vortex configuration never
forms. For both Pc0.5 and Pc1 the $m=2$ (two-vortex configuration)
amplitude decreases from $t\sim 50P_0$. For case Pc1, the vortices are
transient features and are entirely absent for $t\gtrsim80P_0$.

\subsubsection{Strictly isothermal discs} 
\begin{figure}
  \centering
  \includegraphics[width=\linewidth]{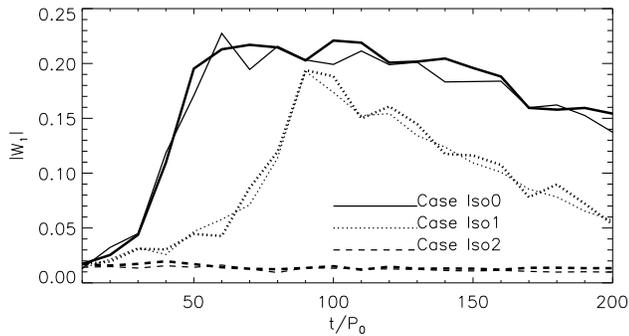}
  \caption{Same as Fig. \ref{pdisk_kerz_cases_planet} but for strictly
    isothermal Iso0 (solid, no viscous layer so $\hat{\nu}\sim10^{-6}$), Iso1
    (dotted, viscous layer with $\hat{\nu}\sim10^{-5}$) and Iso2
    (dashed, viscous layer with $\hat{\nu}\sim10^{-4}$). The thick (thin) lines indicate
    $W_1$ averaged over $\tan{\psi}\in[0,2]h$
    ($\tan{\psi}\in[2,3]h$).  
    \label{pdisk_kerz_cases_planet_iso}}
\end{figure}

We repeated simulations Pb0, Pb1 and Pc1 with a strictly isothermal
equation of state ($q=0$). These are summarised in Table
\ref{planet_sims_iso}. Fig. \ref{pdisk_kerz_cases_planet_iso} compares
their $m=1$ kinetic energy density evolution at the outer gap edge. Consistent with the
above simulations, a viscous layer causes a faster decay in this
quantity. Most interesting though, is that we found case Iso2 (with a viscous
layer of $\sim H$) only shows very weak non-axisymmetric perturbations
early on ($t\lesssim 50P_0$): vortex formation is suppressed.


\begin{table}
  \centering
  \caption{Disc-planet simulations with a strictly
    isothermal equation of state ($q=0$). The thickness of the viscous layer
    is quoted at the reference radius $R=r_0$. \label{planet_sims_iso}}
    \begin{tabular}{llllrc}
      \hline\hline
      Case & $10^6\hat{\nu}_0$ & $A_\nu$ & visc. layer& $10^2\overline{a}_1$ &vortex \\ 
      \hline
      Iso0   & 1.0  & 1          & 0      & 19.3  &  YES   \\
      Iso1   & 1.0  & 10         & $H_0$  & 12.8  &  YES     \\ 
      Iso2   & 1.0  & 100        & $H_0$  & 1.1  &  NO     \\ 
      \hline
  \end{tabular}
\end{table}

\section{Summary and discussion}\label{summary}
We have performed customised hydrodynamic simulations of
non-axisymmetric instabilities in 3D viscous discs. We adopted
height-dependent kinematic viscosity profiles, such that the disc
midplane is of low viscosity ($\alpha\sim 10^{-4}$) and the disc
atmosphere is of high viscosity ($\alpha\sim 10^{-2}$). We were motivated  
by the question of whether or not the Rossby wave instability, and
subsequent vortex formation, operates in layered accretion discs.  

We first considered viscous disc equilibria with a radial density
bump and varied the vertical dependence of viscosity. 
This setup can isolate the effect of viscosity on the
linear RWI. We found that the linear RWI is unaffected by viscosity,
layered or not. The viscous RWI remains dynamical and leads to vortex
formation on timescales of a few 10s of orbits. We continued these
simulations into the non-linear regime, but found that vortices became
stronger as the viscous layer is increased in thickness. We suggest this
counter-intuitive result is an artifact of the chosen viscosity
profile because it is radially structured: viscosity attempts to
restore the equilibrium radial density bump, which favours the
RWI. This effect outweighs viscosity damping the linear instability. 

We also simulated vortex formation at planetary gap edges in layered
discs with a radially-smooth viscosity profile. 
Although vortex formation still 
occurs in layered discs, we found the vortex can be destroyed even when the viscous
layer only occupies the uppermost scale-height of the vertical domain
which is 3 scale-heights. This is significant because most of the disc
mass is contained within 2 scale-heights (i.e. the low viscosity
layer) but simulations show a viscous atmosphere inhibits long term
vortex survival.    
We found that the non-axisymmetric energy densities have
weak vertical dependence, so the disturbance evolves
two-dimensionally. It appears that applying a large viscosity in the
disc atmosphere is sufficient to damp the instability throughout the
vertical column of the fluid.

\cite{barranco05} have described two 
3D vortex models: tall columnar vortices and short finite-height vortices. 
Rossby vortices are columnar, i.e. the associated vortex lines
extend vertically throughout the fluid column. One might have expected
an upper viscous layer to damp out vortex motion in the disc atmosphere, leading to a shorter
vortex. This, however, requires vortex lines to loop around the
vortex (the short vortex of \citeauthor{barranco05}). Such  
vortex loops form the surface of a torus \citep[see, for 
  example, Fig. 1 in][]{barranco05}, instead of ending on 
vertical boundaries. This implies significant vertical
motion near the vertical boundaries of the vortex, which would be  
difficult in our model because of viscous damping applied there. We
suspect this is why short/tall vortices fail to form/survive in our
layered disc-planet models. We conclude that vortex
survival at planetary gap edges require low viscosity
($\alpha\lesssim10^{-4}$) throughout the vertical extent of the
disc.     


\subsection{Relation to other works}
\cite{pierens10} simulated the orbital migration of giant planets in
layered discs by prescribing a height-dependent viscosity profile. They
considered significant reduction in kinematic viscosity in going from
the disc atmosphere (the active zone, with $\alpha\sim10^{-2}$) to the
disc midplane (the dead zone, with $\alpha\sim10^{-7}$). According to
previous 2D simulations, such a low kinematic viscosity should lead to
the RWI \citep{valborro06,valborro07}. However, \cite{pierens10}
did not report vortex formation, nor are vortices visible from their
plots.  Very recent MHD simulations of giant planets in a layered
  disc also did not yield vortex formation \citep{gressel13}. 
These results are consistent with our simulations. 

\cite{oishi09} carried out MHD shearing box simulations with 
a resistivity profile that varied with height to model a
layered disc: the disc atmosphere was MHD turbulent while the disc
midplane remained stable against the MRI. They envisioned the active
zone as a vorticity source for vortex formation in the midplane.  
Although their setup is fundamentally different to ours, they also
reported a lack of coherent vortices in the dead zone. They argued
that the MHD turbulence in the active layer was not sufficiently strong
to induce vortex formation in the dead zone. If MHD turbulence can be
represented by a viscosity, the lack of tall columnar vortices in
\cite{oishi09} is consistent with our results. That is,
even when MRI turbulence is only present in the disc atmosphere it is
able to damp out columnar vortices.     

\subsection{Caveats and outlooks}\label{caveats}

The most important caveat of the current model is the viscous
prescription to mimic MRI turbulence. In doing so, an implicit
averaging is assumed \citep{balbus99}. The spatial averaging should be
taken on length scales no less than the local disc scale-height, and
the temporal average taken on timescales no less than the local
orbital period. These are, however, the relevant scales for vortex
formation via the RWI. Furthermore, our viscosity profile varies on
length-scales comparable to or even less than $H$ (e.g. the vertical transition
between high and low viscosity layers). Nevertheless, our simulations demonstrate 
the importance of disc vertical structure on the RWI. That is,
damping, even confined to the disc atmosphere, can destroy Rossby
vortices.   

Another drawback of a hydrodynamic viscous disc model is the
fact that it cannot mimic magneto-elliptic instabilities (MEI), which
are known to destroy vortices in magnetic discs
\citep{lyra11,mizerski12}. A natural question is how 
Rossby vortices are affected by the MEI when it only operates in
the disc atmosphere. Extension of the present work to global non-ideal
MHD simulations will be necessary to address RWI vortex formation in
layered discs.   

However, some improvements can be made within the viscous
framework. A static viscosity profile neglects the
back-reaction of the density field on the kinematic viscosity. Thus,
our simulations only consider how Rossby vortices 
respond to an externally applied viscous damping. A more
physical viscosity prescription should depend on the local column density
\citep{fleming03}, with viscosity decreasing with increasing column
density. The effective viscosity inside Rossby vortices would be
lowered relative to the background disk because disc vortices are
over-densities. If the over-density is large, then it is
conceivable that vortex formation itself may render the effective
viscosity to be sufficiently low throughout the fluid column to allow
long term vortex survival.    
Preparation for this study is underway and results will be reported in
a follow-up paper.

\section*{Acknowledgments}
This work benefited from extensive discussion with O. Umurhan. I also
thank R. Nelson for discussion, and M. de Val-Borro for a
helpful report.   Computations were performed on the
CITA Sunnyvale cluster, as well as the GPC supercomputer at the SciNet
HPC Consortium. SciNet is funded by the Canada Foundation for
Innovation under the auspices of Compute Canada, the Government of
Ontario, Ontario Research Fund – Research Excellence and the
University of Toronto.   


\appendix
  \section{Artificial radial density bumps with a radially smooth
    viscosity profile}\label{add_sim}
  In \S\ref{density_bump} we found that vortices became stronger
  as the viscous layer thickness is increased, even though linear growth rates
  were reduced. Here, we present additional simulations to support the 
  hypothesis that this is due to the localised radial structure in the viscosity profile. 

\begin{figure}
  \centering
  \includegraphics[width=\linewidth]{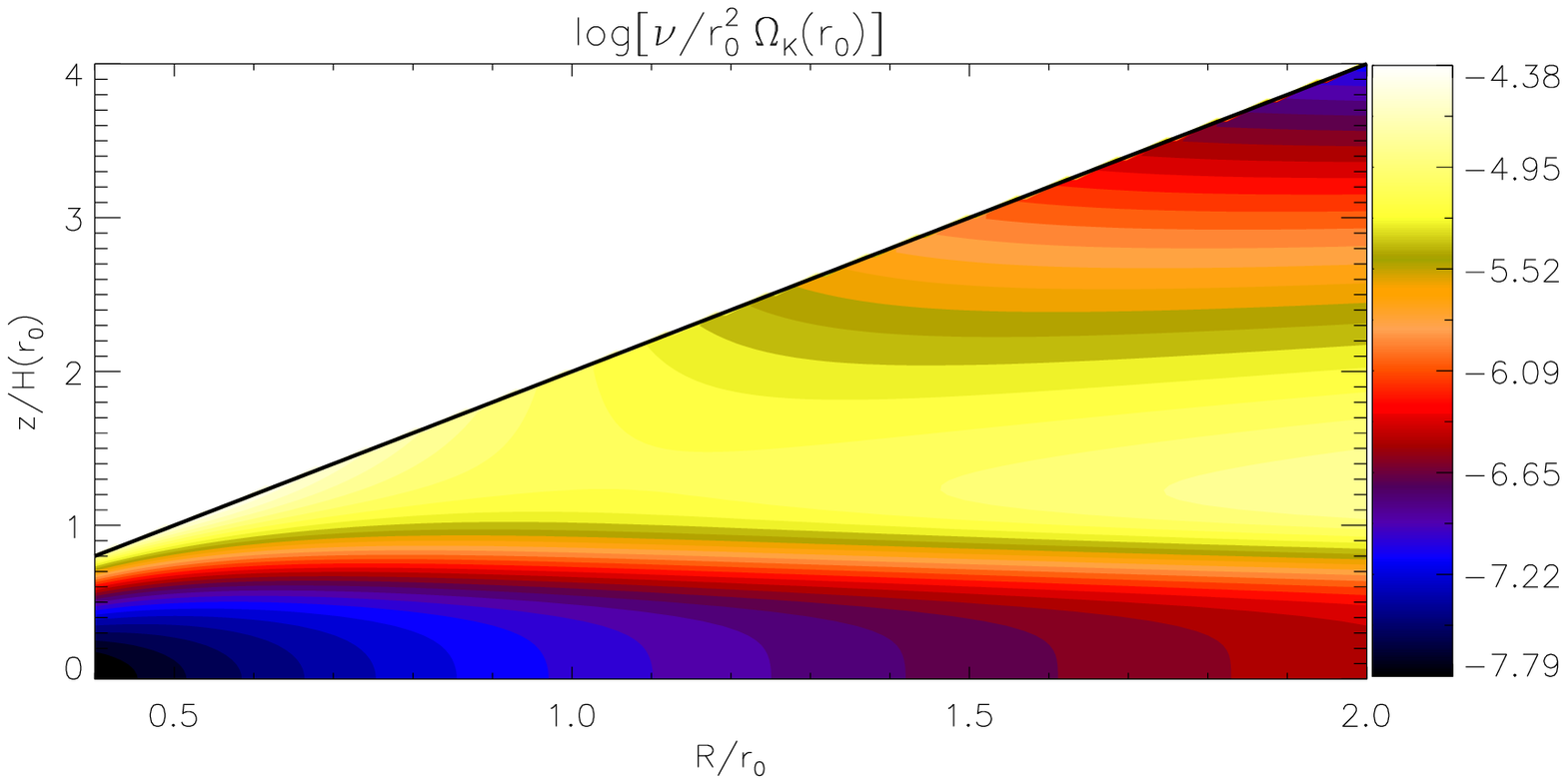}
  \caption{The radially-smooth viscosity profile given by Eq. \ref{smooth_visc}.
         This plot is to be compared with Fig. \ref{visc2d}.\label{appen0}}
  \end{figure}

  We repeated simulation V2 (see Table \ref{artificial_bump})
  with a radially-smooth viscosity profile given by 
  \begin{align}\label{smooth_visc}          
    \hat{\nu}\frac{\rho_i(R,z)}{B(R)} =
    \hat{\nu}_0\left[1+Q(z/H_0) \right]\frac{\rho_i(r_0,z)}{B(r_0)}. 
  \end{align}                  
   Recall the functions $B$ and $Q$ are given by Eq. \ref{bump_func} 
  and \ref{step}, respectively. 
  We set the floor viscosity $\hat{\nu}_0=10^{-7}$ to mitigate
  axisymmetric viscous diffusion of the initial density bump. The
  viscous layer with $\hat{\nu} \sim 10^{-5}$ occupies $z\in[1,2]H_0$
  at $R=r_0$. This viscosity profile is shown in Fig. \ref{appen0}. 
                      
\begin{figure}
  \centering
  \includegraphics[width=\linewidth]{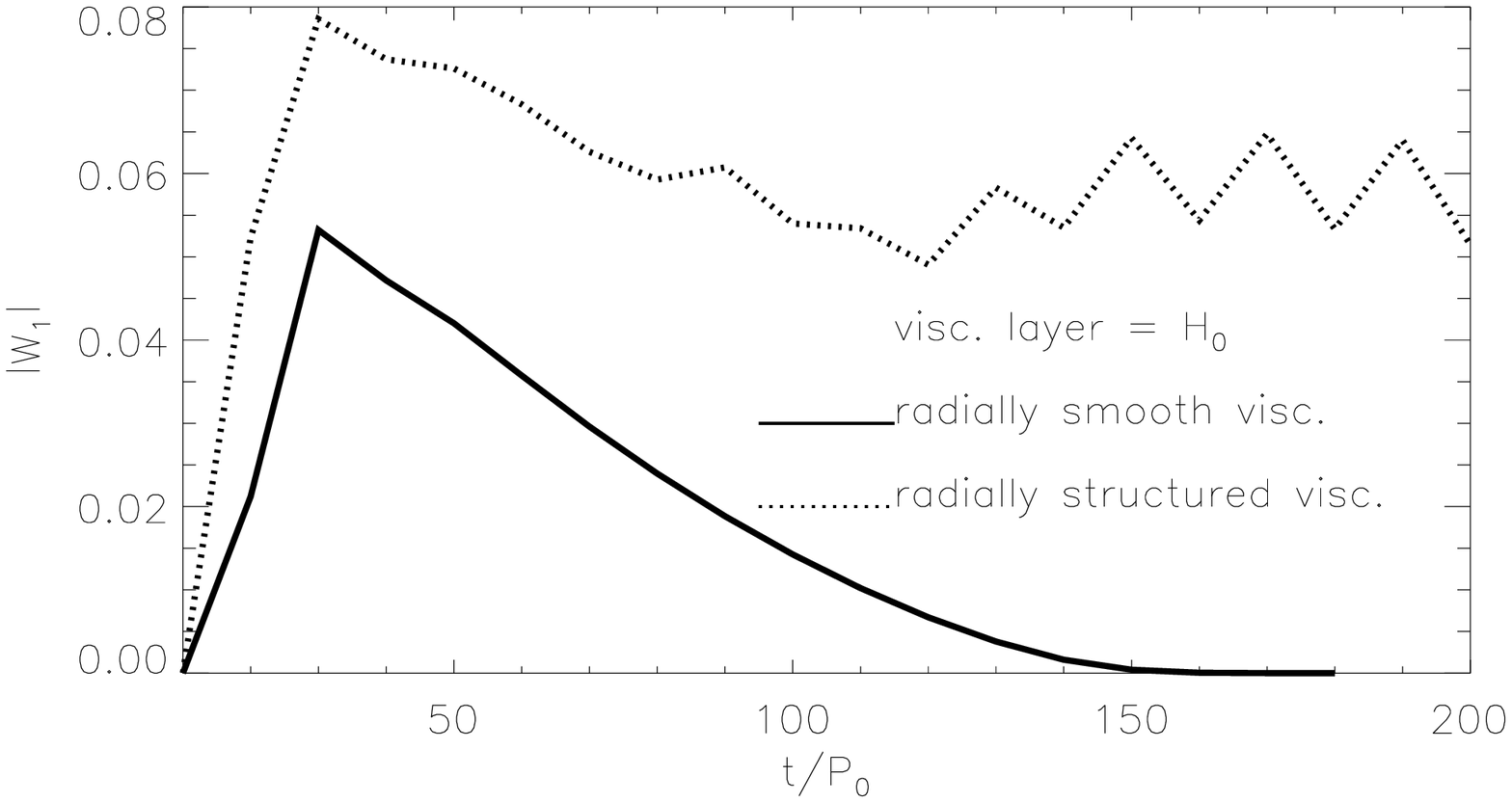}
  \caption{Evolution of the $m=1$ component of the kinetic energy
    density, averaged over the shell $r\in[0.8,1.2]r_0$, for a layered 
    disc initialised with a radial density bump. 
    The solid line employs the radially-structured viscosity profile given by 
    Eq. \ref{visc_profile} (see Fig.\ref{visc2d}). The dotted line employs the radially-smooth
    viscosity profile given by Eq. \ref{smooth_visc} and shown in Fig. \ref{appen0}. \label{appen} }
  \end{figure}

  This simulation is shown as the dotted line in Fig. \ref{appen} in
  terms of the $m=1$ component of the kinetic energy density. We
  compare it to the corresponding case using the radially-structured
  viscosity profile in \S\ref{density_bump} (i.e. the original case V2 but with
  lowered floor viscosity). Vortex formation occurs
  in both runs.  
  With a radially-smooth viscosity profile, the vortex decays 
  monotonically after $|W_1|$ reaches maximum value of $\sim 0.05$. 
  Using the radially-structured viscosity profile (solid line) gives a
  larger disturbance amplitude at the linear stage  ($\max{|W_1|}\sim
  0.08$), and although it subsequently decays, the decay is halted for
  $t\gtrsim110P_0$.    

  The contrast between these cases show that the radial structure
  in the viscosity profile helps vortex survival. This experiment
  indicates that the dominant effect of viscosity is its
  influence on the evolution of the axisymmetric part of background
  disc. The radially-structured viscosity profile is a source for the
  radial PV minimum, which is needed for the RWI.  

  Our result here is qualitatively similar that in \cite{regaly12},
  where a sharp viscosity profile was imposed in a 2D simulation and 
  vortex formation ensues via the RWI. The vortex eventually
  disappears, but re-develops after the system returns to  an
  axisymmetric state. This is because the imposed viscosity profile
  causes the disc to develop the required PV minimum for the RWI. 

\end{document}